\documentclass[a4paper,UKenglish,cleveref, autoref, thm-restate]{lipics-v2021}

\hideLIPIcs  %

\usepackage{etoolbox} %

\usepackage{cite}
\usepackage{amsmath,amssymb,amsfonts}
\usepackage{graphicx}
\usepackage{textcomp}
\usepackage{xcolor}

\newcommand{\algo}[1]{\hbox{\textsc{#1}}}

\usepackage{booktabs}      %
\usepackage{multirow}      %

\usepackage{algorithm}
\usepackage{algpseudocode}
\algdef{SE}[PAR]{ParallelFor}{EndParallelFor}[1]%
{\textbf{for} #1 \algorithmicdo \textbf{ in parallel}}%
{\algorithmicend\ \textbf{parallel for}}
\algdef{SE}[PAR]{ParallelReduce}{EndParallelReduce}[1]%
{\textbf{for} #1 \algorithmicdo \textbf{ reduce in parallel}}%
{\algorithmicend\ \textbf{parallel reduce}}
\algtext*{EndWhile}%
\algtext*{EndIf}%
\algtext*{EndFor}%
\algtext*{EndProcedure}%
\algtext*{EndParallelFor}%
\algtext*{EndParallelReduce}%

\usepackage{float}
\usepackage{todonotes}
\usepackage{dirtytalk}
\usepackage{footnote}
\usepackage{balance}
\usepackage{mathtools}
\usepackage{hyperref}
\usepackage{siunitx}
\title{GPU-Accelerated Algorithms for Process Mapping}

\author{Petr Samoldekin}{Heidelberg University, Heidelberg, Germany}{}{}{}
\author{Christian Schulz}{Heidelberg University, Heidelberg, Germany}{christian.schulz@informatik.uni-heidelberg.de}{https://orcid.org/0000-0002-2823-3506}{}
\author{Henning Woydt\footnote{corresponding author}}{Heidelberg University, Heidelberg, Germany}{henning.woydt@informatik.uni-heidelberg.de}{https://orcid.org/0009-0004-2234-2869}{Funded by the Deutsche Forschungsgemeinschaft (DFG, German Research Foundation) -- DFG SCHU 2567/6-1}

\authorrunning{P. Samoldekin, C. Schulz and H. Woydt}

\Copyright{Petr Samodelkin, Christian Schulz and Henning Woydt}

\ccsdesc[500]{Mathematics of computing~Solvers}
\ccsdesc[500]{Mathematics of computing~Graph algorithms}

\keywords{GPU, Process Mapping, Graph Partitioning}

\nolinenumbers

\begin{document}

    \maketitle

    \begin{abstract}
        Process mapping asks to assign vertices of a task graph to processing elements of a supercomputer such that the computational workload is balanced while the communication cost is minimized.
        Motivated by the recent success of GPU-based graph partitioners, we propose two GPU-accelerated algorithms for this optimization problem.
        The first algorithm employs hierarchical multisection, which partitions the task graph alongside the hierarchy of the supercomputer.
        The method utilizes GPU-based graph partitioners to accelerate the mapping process.
        The second algorithm integrates process mapping directly into the modern multilevel graph partitioning pipeline.
        Vital phases like coarsening and refinement are accelerated by exploiting the parallelism of GPUs.
        The first algorithm has, on average, about 12 percent higher communication costs than the state-of-the-art solver and thus remains competitive with it.
        However, in terms of speed, it vastly outperforms the competitor with a geometric mean speedup of 22 times and a maximum speedup of 934 times.
        The second approach is even faster, with a geometric mean speedup of 1454 times and a peak speedup of 12376 times.
        Compared to other algorithms that prioritize speed over solution quality, this approach has the same quality but much greater speedups.
        To our knowledge, these are the first GPU-based algorithms for process mapping.
    \end{abstract}

    \section{Introduction}\label{sec:introduction}
    Parallel high-performance applications rely on the efficient execution of large numbers of tasks across massively parallel architectures.
    This involves placing individual tasks on the Processing Elements (PEs) of the underlying hardware architecture such that the computational workload is evenly balanced.
    However, load balancing alone is not sufficient.
    In parallel applications, the individual tasks usually communicate with each other, i.e., they send and receive data to and from each other.
    The cost of this communication heavily depends on the location of the tasks in the hardware topology.
    Modern supercomputers are typically organized hierarchically, for example into processors, nodes, racks, and islands.
    The communication cost between two PEs, therefore, depends on their relative locations within this hierarchy.
    Two PEs on the same processor communicate data much more quickly than two PEs that are located on different islands.
    Thus, in addition to balancing the workload, it is advantageous to minimize the total communication cost by placing highly interactive tasks close to each other.
    This optimization problem is known as the \emph{general process mapping problem} (GPMP).

    Optimally solving the GPMP is $\mathrm{NP}$-hard~\cite{Sahni76} as it involves solving the well-known \emph{quadratic assignment problem} (QAP)~\cite{Burkard98}.
    The QAP considers a set of $n$ facilities and $n$ locations.
    The objective is to assign each facility to exactly one location in a way that minimizes the total cost.
    This cost is determined by the flow between pairs of facilities and the distance between the locations to which they are assigned.
    In other words, facilities with higher interaction should ideally be placed closer together to reduce the overall cost.
    Since the QAP is strongly $\mathrm{NP}$-hard, i.e., no constant factor approximation exists unless $\mathrm{P} = \mathrm{NP}$, its hardness transfers to the GPMP\@.
    The hardness is also reflected in practice as no meaningful instance with $n > 20$ can be solved in a reasonable time~\cite{Burkard98}.
    Moreover, the same hardness result arises from \emph{graph partitioning} (GP).
    GP considers a graph $G=(V, E)$ and asks for a partition of the vertex set into $k$ pairwise disjoint sets of equal size, such that the number of edges crossing between the different subsets is minimized.
    Graph partitioning is $\mathrm{NP}$-complete~\cite{Garey76} and also admits no constant-factor approximation unless $\mathrm{P} = \mathrm{NP}$~\cite{Andreev06}.

    The GPMP is commonly solved using either the \emph{two-phase} approach or via \emph{integrated mapping}.
    In the two-phase approach, the task set is first partitioned into as many (almost) equally sized sets as there are PEs in the supercomputer.
    While partitioning the task set, the necessary volume of communication across the different resulting sets is minimized.
    In the second phase, the subsets are mapped to the PEs such that the total cost of communication is minimized, taking the distances between the PEs into account.
    While the first phase is an instance of graph partitioning, the second phase is an instance of the quadratic assignment problem.
    This makes the approach easier to realize as one can fall back to two well-studied problems.
    The algorithm \algo{SharedMap}~\cite{Schulz25} currently offers state-of-the-art solution quality by utilizing \emph{hierarchical multisection}.
    This special variant of the two-phase approach directly exploits the hardware topology of the supercomputer.
    The integrated mapping approach combines both phases into a single process.
    The mapping objective of the second phase is directly integrated into the partitioning of the first phase.
    Rather than minimizing the communication volume between the subsets, it directly minimizes the total cost of communication.
    Typically, this is achieved using the modern multilevel graph partitioning framework.
    The input graph is first coarsened to a smaller representation and then an initial partition is determined.
    During uncoarsening, the partition is refined using local search algorithms.
    The approach is more complex as the whole graph partitioning pipeline has to be adopted to the new objective function.
    However, this also enables function-specific optimization resulting in an overall faster running time.
    Integrated mapping algorithms, like \algo{IntMap}~\cite{Faraj20}, often achieve faster running times compared to the two-phase approach.

    Recently, Graphics Processing Units (GPUs) have been utilized to tackle graph partitioning.
    GPUs provide many more cores than CPUs and, when exploited efficiently, can significantly reduce the running time.
    However, effectively leveraging this level of parallelism for graph algorithms is challenging.
    Graph computations typically involve irregular memory access patterns, uneven workloads across vertices, and complex data dependencies.
    Despite this, recent GPU-based algorithms like \algo{Jet}~\cite{Gilbert24} and \algo{G-kway}~\cite{Lee24} have demonstrated promising results.
    They deliver solution qualities comparable to the serial CPU-based graph partitioner \algo{Metis}~\cite{Karypis98} while being up to 20 times faster.
    Both the two-phase and integrated mapping approach rely on graph partitioning, so these results should transfer to GPMP\@.

    \textbf{Our Contribution.}
    We present two GPU-based algorithms to solve the process mapping problem.
    The first one is based on the two-phase approach and utilizes hierarchical multisection.
    In contrast, the second algorithm integrates the mapping into the graph partitioning pipeline.
    Both significantly outperform state-of-the-art CPU solvers in runtime, achieving speedups of more than $900\times$, while remaining competitive with their solution quality.

    \section{Concepts and Notations}\label{sec:concepts-and-notations}
    Let $G=(V, E)$ with $n = |V|$ and $m = |E|$ be an undirected graph, and let $c : V \mapsto \mathbb{R}$ and $\omega : E \mapsto \mathbb{R}$ be vertex and edge weights.
    Their natural extension to sets is $c(V') = \sum_{v \in V'} c(v)$ and $\omega(E') = \sum_{e \in E'} \omega(e)$ for vertex sets $V' \subseteq V$ and edge sets $E' \subseteq E$.

    The \emph{graph partitioning problem} (GPP) asks for a partition of the vertex set $V$ into $k$ equally sized sets.
    More formally, let $k \in \mathbb{N}$ be the desired number of blocks and let $\epsilon \geq 0$ be a hyper-parameter called \emph{imbalance}.
    GPP then asks for a partition $V = V_1 \cup \ldots \cup V_k$ such that $\forall i \neq j : V_i \cap V_j = \emptyset$ and each $V_i$ adheres to the balance constraint, that is, \hbox{$c(V_i) \leq L_{\max} \coloneqq \lceil (1 + \epsilon)\frac{c(V)}{k} \rceil$}.
    Such a partition is called a \emph{$k$-way partition} of $G$.
    We call a block $V_i$ \emph{overloaded} if $c(V_i) > L_{\max}$ and \emph{underloaded} if $c(V_i) \leq L_{\max}$.
    Let \hbox{$E_{ij} = \{\{u, v\} \in E \mid u \in V_i, v \in V_j \}$} be the edges crossing between block $V_i$ and $V_j$.
    The \emph{edge-cut} of a partition is the total weight of edges crossing between blocks, i.e., $\sum_{i < j} \omega(E_{ij})$.
    The goal of GP is to determine a balanced $k$-way partition that minimizes the edge-cut.

    A parallel application consists of $n$ weighted tasks which communicate with each other.
    This is captured by the communication matrix $\mathcal{C} \in \mathbb{R}^{n \times n}$, where an entry $\mathcal{C}_{ij}$ denotes the communication volume between tasks $i$ and $j$.
    An entry $\mathcal{C}_{ij} = 0$ denotes that the tasks do not communicate with each other.
    Most modern parallel applications only have a very sparse communication pattern, which is why the application can also be modeled as a task graph $G_{\mathcal{C}}$.
    For each task in the application there exists one vertex with the same weight.
    For each non-zero entry $\mathcal{C}_{ij}$, the task graph has one forward and backward edge with weight $\mathcal{C}_{ij}$.
    It is assumed that the communication is symmetric, otherwise it can be modeled symmetrically~\cite{Brandfass12}.

    Let $k$ be the number of PEs on a supercomputer.
    The cost of communication between two PEs is captured by the hardware topology matrix $\mathcal{D} \in \mathbb{R}^{k \times k}$.
    An entry $\mathcal{D}_{xy}$ denotes a cost factor that is applied to communication between PEs $x$ and $y$.
    The cost of tasks $i$ and $j$ communicating, if placed on PEs $x$ and $y$ respectively, is therefore $\mathcal{C}_{ij}\mathcal{D}_{xy}$.
    It is assumed that all PEs can communicate with each other and that the matrix is symmetric.

    The \emph{general process mapping problem} (GPMP) is then defined as follows.
    Let $G_{\mathcal{C}} = (V, E)$ with $n = |V|$ be a task graph, let $\mathcal{D}$ be the hardware topology matrix of a supercomputer with $k$ PEs and let $\epsilon$ be the allowed imbalance.
    GPMP then asks to partition the task set $V$ into $k$ equally sized subsets, such that the overall communication cost is minimized.
    More formally, determine a mapping $\Pi : [n] \mapsto [k]$ such that $J(\mathcal{C}, \mathcal{D}, \Pi) \coloneqq \sum_{i, j}\mathcal{C}_{ij}\mathcal{D}_{\Pi(i)\Pi(j)}$ is minimized, while each block adheres to the balance constraint.

    In the \emph{hierarchical process mapping problem} (HPMP), a special variant of the GPMP, the hardware topology of the supercomputer is described by a \emph{hierarchy}~\hbox{$H = a_1:\ldots:a_\ell$}.
    It denotes that each processor has $a_1$ PEs, each node has $a_2$ processors, each rack has $a_3$ nodes, and so forth.
    In total, the system has $k = \prod_{i \leq \ell} a_i$ PEs.
    Corresponding to the hierarchy, a distance $D = d_1:\ldots:d_\ell$ is given which denotes the cost factor between the PEs.
    Two PEs on the same processor have a cost factor of $d_1$, two PEs on the same node but on different processors have a cost factor of $d_2$, two PEs on the same rack but on different islands have a cost factor of $d_3$, and so forth.
    The focus of this work lies in solving the HPMP\@.

    In GPMP, $n > k$ is typically assumed, i.e., there are (considerably) more tasks than PEs.
    If $n = k$, the problem is known as the \emph{one-to-one process mapping problem} (OPMP) which is identical to the \emph{quadratic assignment problem} (QAP).
    The QAP is commonly solved using refactorization and linearization techniques~\cite{Adams93, Adams07, Hahn12} of its corresponding mixed integer linear programming formulation~\cite{Lawler63}.
    We refer the reader to~\cite{Burkard98, Anstreicher03, Loiola07} for comprehensive surveys.
    The problems GPP, GPMP, HPMP, and QAP are all NP-hard problems~\cite{Garey76, Sahni76} and no constant factor approximation guarantee exists \hbox{unless P = NP}.

    \subsection{Multilevel Graph Partitioning}\label{subsec:multilevel-graph-partitioning}
    Modern graph partitioning relies on the multilevel approach, which works in three phases: coarsening, partitioning, and uncoarsening with refinement.
    In the \emph{coarsening} phase, a succession of smaller graphs $G, G_1, \ldots, G_t$ is computed, with the goal that a smaller graph $G_{i+1}$ retains the same overall structure as $G_i$.
    A \say{good} partition of $G_{i+1}$ would then also resemble a \say{good} partition for $G_i$.
    Coarsening is applied until the coarsest graph has fewer than $ck$ vertices, where $c$ is a tunable constant.
    This small graph can then be efficiently partitioned.
    The contractions are then undone and the quality of the partition is improved.

    To \emph{coarsen} a graph, one usually applies a matching or a clustering algorithm.
    The matching algorithm rates each edge and then determines a maximum weighted matching.
    The edges selected in the matching are then contracted.
    Contracting an edge $\{u, v\}$ replaces the vertices $u$ and $v$ with a new vertex $w$ whose weight is defined as $c(w) = c(u) + c(v)$.
    All former neighbors of $u$ and $v$ are subsequently connected to $w$.
    Parallel edges are fused and their weights are summed.
    A clustering algorithm first determines clusters in the graph, and then contracts each cluster into one super-vertex.
    The vertex weights and parallel edges are handled in the same way as in the matching.

    In the \emph{initial partitioning phase}, the coarsest graph $G_t$ is partitioned.
    Since $G_t$ is very small, expensive but high-quality algorithms, such as recursive bisection or direct $k$-way partitioning, can be applied to minimize the edge-cut.

    The \emph{uncoarsening and refinement} phase iteratively uncontracts the graphs $G_{i+1}$ into $G_i$ by undoing the contractions.
    After each uncontraction, local search algorithms are applied to improve the objective.
    The algorithms try to move vertices across the blocks to minimize the edge-cut, while respecting the balance constraint.
    Widely used refinement algorithms are the Fiduccia-Mattheyses (FM) algorithm~\cite{Fiduccia82}, label propagation~\cite{Raghavan07}, and flow-based refinement~\cite{Sanders11}.
    Notable serial graph partitioners are \algo{Metis}~\cite{Karypis98}, \algo{KaFFPa}~\cite{Sanders11}, \algo{Jostle}~\cite{Walshaw00}, \algo{Scotch}~\cite{Pellegrini96} and \algo{KaHyPar}~\cite{Schlag20}.

    \section{Related Work}\label{sec:related-work}
    Process mapping is closely related to graph partitioning, which has already seen a tremendous amount of research over the last few decades.
    We already gave a brief overview in Section~\ref{subsec:multilevel-graph-partitioning} and refer the reader to~\cite{Bichot13, Catalyurek23, Buluc16, Schloegel03} for more comprehensive surveys.
    We will give here an overview of \algo{Jet}~\cite{Gilbert24}, a GPU-based graph partitioner, and of process mapping.

    \subsection{GPU Multilevel Graph Partitioning}\label{subsec:gpu-multilevel-graph-partitioning}
    Gilbert~et~al.~\cite{Gilbert24} present \algo{Jet}, a GPU-accelerated multilevel graph partitioner designed to achieve high partition quality and fast runtimes.
    Their main contribution includes a novel GPU-parallel refinement algorithm based on label propagation~\cite{Raghavan07}.
    They implemented their algorithm in the performance-portable Kokkos framework~\cite{Edwards14}.

    \textbf{Matching.}
    \algo{Jet} utilizes a two-hop matching~\cite{LaSalle15}, originally developed for \algo{Mt-Metis}~\cite{Lasalle13}, a shared-memory graph partitioner.
    The algorithm first applies a heavy-edge matching and only if more than 25\% of vertices remain unmatched, a two-hop matching is applied.
    A heavy-edge matching determines for each vertex its heaviest edge and two neighbors are matched if they agree on their heaviest edge.
    The two-hop matching can be split into three categories: \emph{leaf}, \emph{twin} and \emph{relative} vertices.
    Leaf vertices have degree one, two twin vertices share the same neighborhood and relative vertices share at least one common neighbor.
    To identify twin vertices the neighborhoods of all vertices are hashed and sorted, so vertices with the same neighborhood can easily be found.
    To accelerate the matching of relatives, special matchmaker vertices with small degree are used.
    At first leaf vertices that share a neighbor are matched, then twin vertices and at the end, relative vertices, if necessary.

    \textbf{Coarsening.}
    To coarsen the graph, a fine-grained per-vertex hashing algorithm is deployed.
    First, the maximum neighborhood size of each coarse vertex is overestimated.
    Then a prefix-sum over the sizes is determined, assigning each vertex a subarray.
    Each vertex hashes its new neighboring vertices and inserts them and their weights into their respective subarray.
    In the end, the edges of all vertices are extracted from their hash array.

    \textbf{Initial Partitioning.}
    After coarsening to $4k$–$8k$ vertices, the partitioning of the coarsest graph is delegated to \algo{Metis}~\cite{Karypis98} on the CPU\@.
    The transfer overhead between CPU and GPU is negligible due to the small size of the coarsest graph.

    \textbf{Refinement.}
    The novel improvement of \algo{Jet} is its refinement algorithm, which alternates between a highly parallelized label propagation~\cite{Raghavan07} algorithm and two rebalancing strategies.
    A standard serial label propagation algorithm visits the vertices in random order.
    A vertex is moved to a neighboring block if it improves the edge-cut and adheres to the balance constraint.
    This is often repeated in multiple rounds to further refine the partition.
    However, the local search can get stuck in local optima as only positive moves are allowed.
    Additional problems arise with parallelization.
    Moving multiple vertices at once makes it difficult to maintain the balance constraint.
    Additionally, while each individual vertex move might have a positive impact, moving adjacent vertices in parallel can worsen the solution quality.

    Gilbert~et~al.~\cite{Gilbert24} tackle these problems by splitting the refinement into two parts, first unconstrained label propagation and afterwards rebalancing.
    The unconstrained label propagation starts by determining for each vertex $v$ the gain $F(v)$ (decrease in edge-cut) of moving it into an adjacent block.
    All potential vertex moves are then passed through the first filter $F(v) \geq 0 \lor -F(v) < \lfloor c \cdot \operatorname{conn}(v, \Pi(v))  \rfloor$.
    Here, $\operatorname{conn}(v, b)$ is the sum of edge weights from vertex $v$ to block $b$.
    To pass the first filter a move needs to be non-negative (first condition) or its increase in edge-cut must be comparatively small (second condition).
    The constant $c \in [0, 1]$ controls the number of negative gain moves, with larger values allowing more moves.
    These negative moves are allowed to escape local minima.
    All moves that pass the first filter are collected in a list $\mathcal{X}$, which is implicitly sorted in decreasing order by gain.
    For each vertex $v \in \mathcal{X}$, the algorithm then recomputes the gain~$\mathbb{F}(v)$ under an approximate future partition state: if a neighbor $u$ appears earlier in the ordering, it is assumed to have already moved to its new block.
    In a second filter, only the vertices $v \in \mathcal{X}$ with a non-negative gain~$\mathbb{F}(v)$ are moved to their respective destination.
    To prevent oscillation each vertex that was moved is locked in the next iteration of unconstrained label propagation.

    \textbf{Rebalancing.}
    Since the unconstrained label propagation phase ignores the balance constraint, an additional rebalancing phase is required.
    \algo{Jet} implements two similar strategies, weak and strong rebalancing.
    Weak rebalancing determines for each vertex in an overloaded block $A$ the loss of moving it to a valid destination block.
    Valid destination blocks are blocks $B$ with $c(B) \leq \sigma \leq L_{\max}$.
    The upper bound $\sigma$ is used instead of $L_{\max}$ to create a small buffer zone, so a block is not overloaded after only one move.
    For each overloaded block $A$ its vertices are approximately sorted increasingly by loss in a list $L_A$.
    Then a prefix $L_A^* \subseteq L_A$ is determined such that moving all vertices in $L_A^*$ will balance block $A$.
    This minimizes the loss to balance a specific block, however, the destination blocks might become overloaded, so another iteration might be necessary.
    Strong rebalancing takes extra care by preventing destination block from becoming overloaded.
    If this would happen, the vertices are moved to another underloaded block.
    However, these blocks must not necessarily be connected to the vertex.
    This results in a much greater loss but the partition will most likely be balanced.

    In total, \algo{Jet} combines its unconstrained label propagation and rebalancing in multiple iterations and keeps the best solution encountered.
    The algorithm offers comparable solution quality to the CPU-based graph partitioner \algo{Metis}~\cite{Karypis98}, while being up to 20 times faster.
    Other GPU graph partitioners are \algo{Goodarzi~et~al.}~\cite{Goodarzi19}, \algo{Sphynx}~\cite{Acer20} and \hbox{\algo{G-kway}}~\cite{Lee24}.

    \subsection{Process Mapping}\label{subsec:process-maping}
    Similar to graph partitioning, process mapping has also seen a tremendous amount of research, and we refer the reader to Hoefler~et~al.~\cite{Hoefler13} and Pellegrini~\cite{Pellegrini13} for broader overviews.
    Among the first to implement process mapping was Hatazaki~\cite{Hatazaki98} who used graph partitioning to map MPI processes onto hardware topologies.
    Träff~\cite{Traff02} and Yu~et~al.~\cite{Yu06} implemented mapping algorithms for special hardware, namely for the NEC SX-series of parallel vector computers and the Blue Gene/L Supercomputer, respectively.

    From a theoretical perspective, it appears straightforward that minimizing $J(\mathcal{C}, \mathcal{D}, \Pi)$ will lead to a more efficient algorithm.
    However, the question remains whether the theoretical improvement translates into practical performance gains.
    Brandfass~et~al.~\cite{Brandfass12} show that optimizing the communication cost indeed leads to performance improvements for the simulation of Computational Fluid Dynamics.
    Manwade~et~al.~\cite{Manwade18} show in a more general context that process mapping algorithms improve the performance of MPI~applications.

    There are two main methods to solve process mapping: the \emph{two-phase} approach and \emph{integrated mapping}.
    The two-phase approach first leverages standard graph partitioning to determine a balanced $k$-way partition with small edge-cut of the task graph~$G_\mathcal{C}$.
    This results in a partition with an overall small communication volume between the blocks.
    The second phase then determines a 1-to-1 mapping of blocks to the PEs that minimizes the communication cost.
    Müller-Merbach~\cite{MullerMerbach70} presented a greedy algorithm to obtain an initial mapping that served as a basis for many future algorithms.
    After the mapping phase, a potential third refinement phase can be used to determine individual vertex moves that decrease the communication cost.
    Heider~\cite{Heider72} introduced a simple swap algorithm that considers all possible $\mathcal{O}(n^2)$ swaps.
    This method was iteratively improved by Brandfass~et~al.~\cite{Brandfass12} by avoiding redundant swaps, and Schulz and Träff~\cite{Schulz17} by utilizing better data structures and restricting the search space.
    Notable algorithms are \algo{GlobalMultisection}~\cite{Kirchbach20} and \algo{SharedMap}~\cite{Schulz25}.

    \textsc{GlobalMultisection}~\cite{Kirchbach20} employs a method called \emph{hierarchical multisection} inspired by \textsc{KaFFPa-Map}~\cite{Schulz17}.
    \textsc{KaFFPa-Map} would first create a balanced $k$-way partition and then create a communication model graph $G_\mathbb{M}$.
    This graph has $k$ vertices and edges connecting them if the corresponding $k$-way partition has edges between the blocks.
    $G_\mathcal{M}$ is then perfectly partitioned alongside the hierarchy $H = a_1 : \ldots : a_{\ell}$, first into an $a_\ell$-way partition, then each of the blocks into an $a_{\ell - 1}$-way partition and so forth.
    This is repeated recursively until $k$ blocks are obtained.
    These represent the $k$ PEs of the system.
    The mapping of blocks to PEs, and therefore vertices to PEs, can be determined by following the recursive calls.
    \textsc{GlobalMultisection} improves the approach by not creating the communication model graph $G_\mathbb{M}$, but instead directly determining an $a_\ell$-way partition of $G_{\mathcal{C}}$ and then recursing on each block.
    Since no perfectly balanced partition is needed, \algo{GlobalMultisection} has more freedom to move vertices, resulting in lower communication costs.

    \textsc{SharedMap}~\cite{Schulz25} further decreases the communication cost.
    It introduces an adaptive imbalance strategy that rescales the imbalance parameter $\epsilon$ for each individual partitioning.
    It guarantees that the final $k$-way partition is $\epsilon$-balanced.
    This results in an overall better mapping with lower communication costs.
    \algo{SharedMap} uses the serial graph partitioner \algo{KaFFPa}~\cite{Sanders11} as one of its underlying graph partitioners, and currently offers state-of-the-art solution quality for HPMP\@.

    The second approach to solve the GPMP, \emph{integrated mapping}, combines the partitioning and mapping into one algorithm.
    It leverages the multilevel graph partitioning scheme and the refinement algorithms minimize $J(\mathcal{C}, \mathcal{D}, \Pi)$ instead of edge-cut.
    Two prominent algorithms are \algo{IntMap}~\cite{Faraj20} and \algo{Mt-KaHyPar-Steiner}~\cite{Heuer23}.

    \textsc{IntMap}~\cite{Faraj20} uses matching-based coarsening to create a smaller representation of the graph.
    The Global Path Algorithm~\cite{Maue07}, a $1/2$-approximation algorithm, is used to determine a matching of vertices.
    Instead of using the edge weights as a rating, a special rating function \hbox{$\exp^*(\{u, v\}) = \omega(\{u, v\}) / c(u)c(v)$} from~\cite{Holtgrewe10} is used.
    As initial partitioning, the hierarchical multisection approach is used.
    Refinement strategies include quotient graph refinement~\cite{Sanders11}, $k$-way FM refinement~\cite{Sanders11}, label propagation~\cite{Raghavan07}, and multi-try FM refinement~\cite{Sanders11}.
    All refinement methods use the \emph{gain} of a vertex $v$ when moving it from its current block $\Pi(v)$ to a new block $b$.
    The gain is defined as follows:
    \begin{equation}
        \label{eq:gain}
        G_b(v) = \sum_{u \in N(v)} C_{v, u} \Big(\mathcal{D}_{\Pi(v), \Pi(u)} - \mathcal{D}_{b, \Pi(u)}\Big).
    \end{equation}
    Multiple methods are presented to query the distance $\mathcal{D}_{i, j}$ between two PEs.
    The simplest one saves a matrix with $\mathcal{O}(k^2)$ space and accesses each distance in $O(1)$.
    The other methods, have a considerably lower space complexity, but come with more computational load.

    \textsc{Mt-KaHyPar-Steiner}~\cite{Heuer23} originally minimizes wire lengths in VLSI designs.
    Logical units, a hypergraph~$\mathcal{H}$, are mapped to chip locations, an edge-weighted graph~$\mathcal{C}$, such that the total sum of used edge weights, the implicit wire-length distance, is minimized.
    Replacing $\mathcal{H}$ with the task graph $G_\mathcal{C}$ and $\mathcal{C}$ with the hardware topology matrix~$\mathcal{D}$ leads to GPMP\@.
    \algo{Mt-KaHyPar-Steiner} is integrated into the shared-memory parallel hypergraph partitioner \algo{Mt-KaHyPar}~\cite{Gottesbueren20}.
    The algorithm follows the multilevel (hyper-)graph partitioning approach.
    First, the hypergraph is clustered, and these clusters are contracted into individual vertices, creating a smaller graph.
    An initial $k$-way partition of the hypergraph is determined and then greedily mapped onto the chip locations.
    As local search refinement algorithms, label propagation~\cite{Raghavan07}, FM refinement~\cite{Fiduccia82}, and flow-based refinement~\cite{Sanders11} are used.

    \subsection{Kokkos}\label{subsec:kokkos}
    We utilize Kokkos~\cite{Edwards14}, a C++ library providing performance portability across many-core architectures, including GPUs.
    It abstracts two main concepts: (1) parallel execution patterns and (2) multidimensional arrays.
    Since multidimensional arrays are not used in this work, we focus on the parallel execution patterns, which are built around three fundamental primitives.
    (a) \emph{parallel\_for} executes $N$ independent work units in parallel.
    Each work unit operates on the same user-defined function $f(i)$ with only the index $i \in [0, N)$ changing for each work unit.
    (b) \emph{parallel\_reduce} performs a reduction $R = \bigoplus_{i = 0}^{N-1} f(i)$ over values returned by $f(i)$ using an associative operator $\oplus$.
    (c) \emph{parallel\_scan} computes an exclusive prefix sum over a range of values, i.e. $y_i = \sum_{j = 0}^{i - 1} f(j)$.

    \subsection{CSR Graph Format}\label{subsec:csr-graph-format}
    A graph $G = (V, E)$ is typically stored in the Compressed Sparse Row (CSR) format. %
    The representation consists of an offset array $O$ (also called \emph{row-pointer}) of size $|V|+1$ and two arrays of size $|E|$, one for the edge vertices $E_v$ and one for the edge weights $E_w$.
    The edges incident to vertex $v$ and the corresponding edge weights are stored in the contiguous range $[O[v], O[v + 1])$ of $E_v$ and $E_w$.
    Note that the graph is undirected and an edge $\{u, v\}$ is stored for both $u$ and $v$.
    This layout naturally supports parallelization across vertices.

    \section{GPU-Accelerated Process Mapping}\label{sec:gpu-accelerated-process-mapping}
    The main contribution of this work is the development of two GPU-based approaches for solving the hierarchical process mapping problem.
    Given the close relationship between process mapping and graph partitioning, and the recent success of GPU-based partitioners in achieving faster runtimes, we extend these advantages to process mapping.
    The first algorithm follows the hierarchical multisection approach.
    It recursively partitions the task graph~$G_{\mathcal{C}}$ along the system's hierarchy: first across the islands, then across the racks, and so forth.
    The final mapping of blocks, and thus vertices, to PEs is obtained by following the recursive calls.
    The algorithm is presented in Section~\ref{subsec:hierarchical-multisection}.
    The second algorithm integrates process mapping directly into the multilevel graph partitioning scheme and is inspired by the GPU-based graph partitioner \algo{Jet}~\cite{Gilbert24}.
    Its refinement phase has been adapted to employ the new gain equation (Equation~\ref{eq:gain}).
    Section~\ref{subsec:integrated-mapping} presents the algorithm.

    \textbf{Extended CSR Format.}
    Before presenting the algorithms, we want to present our extended CSR format.
    As introduced in Section~\ref{subsec:csr-graph-format}, the CSR representation naturally supports parallelization over vertices, allowing each thread to process a vertex's adjacency list independently.
    However, some computations benefit from direct parallelization over edges.
    In standard CSR, this requires nested parallel loops, first over vertices and then their neighbors.
    This can lead to poor load balancing due to varying vertex degrees, resulting in idle threads on GPUs.
    To overcome this, we extend the CSR format with an additional array $E_u$ of size $|E|$ that explicitly stores the opposite endpoint of each edge.
    This edge-centric representation enables flat parallelism over all edges, improving load balance and GPU utilization.
    We refer to this edge-list representation as $\mathbb{E}$ throughout the rest of this work.

    \subsection{Hierarchical Multisection}\label{subsec:hierarchical-multisection}
    The hierarchical multisection approach was first introduced by Kirchbach~et~al.~\cite{Kirchbach20}.
    It directly exploits the given hierarchy \hbox{$H=a_1:\ldots:a_\ell$} of the supercomputer to partition the task graph~$G_{\mathcal{C}}$.
    First, an $a_{\ell}$-way partition of $G_{\mathcal{C}}$ is determined utilizing a graph partitioning algorithm.
    The resulting $a_{\ell}$~blocks represent the largest hardware components of the supercomputer (e.g., the islands).
    Since the graph partitioner minimizes the edge-cut between the blocks, it directly corresponds to minimizing the communication cost across the islands.
    Then the algorithm recursively partitions each of the $a_{\ell}$ blocks into an $a_{\ell - 1}$-way partition.
    These blocks represent the second-largest component of the system (e.g., racks within islands).
    Again, minimizing the edge-cut corresponds to minimizing the communication across the racks of each island.
    This recursive procedure is continued until $k$ blocks are obtained.

    Schulz and Woydt~\cite{Schulz25} improved this method by introducing an adaptive imbalance parameter.
    Using the original imbalance parameter $\epsilon$ for each partitioning can result overall in an unbalanced $k$-way partition.
    To avoid this problem, an adaptive imbalance parameter $\epsilon'$ is determined for each partitioning.
    It depends on the original imbalance $\epsilon$, the overall graph weight $c(V)$, the weight of the current subgraph $c(V')$, the desired number of blocks $k$, the number of total blocks the current subgraph will be partitioned into $k'$ and the depth of the current partition in the hierarchy $d$.
    The adaptive imbalance is then determined by
    \begin{equation}
        \label{eq:adaptive-imbalance}
        \epsilon' = \Big( (1+\epsilon) \frac{k'c(V)}{kc(V')} \Big)^{\tfrac{1}{d}} - 1.
    \end{equation}
    Using the adaptive imbalance guarantees that the final partition is \hbox{$\epsilon$-balanced}.
    Experimental results in~\cite{Schulz25} demonstrate that the adaptive imbalance improves the overall mapping quality.
    In particular, \algo{SharedMap}~\cite{Schulz25} currently achieves state-of-the-art quality for the HPMP\@.

    To adopt the hierarchical multisection approach to GPUs, we must (1) employ a GPU-based graph partitioner to obtain the $a_i$-way partitions and (2) construct the resulting subgraphs directly on the GPU\@.
    For the partitioning step, we use our own reimplementation of \algo{Jet}~\cite{Gilbert24}, which we already presented in Section~\ref{subsec:gpu-multilevel-graph-partitioning}.
    \algo{Jet} outperforms the CPU-based graph partitioner \textsc{Metis}~\cite{Karypis98} in terms of speed by a factor in the range of 2 to 20 times while delivering equal or better solution quality.
    Alternative GPU-based graph partitioners such as \algo{Sphynx}~\cite{Acer20} or \algo{G-kway}~\cite{Lee24} could also be used.
    However, no comprehensive comparison between GPU partitioners exists, so we adopt the implementation of \algo{Jet} for its thorough evaluation.
    Any faster or higher-quality graph partitioner would also improve this algorithm.

    For the second step, the induced subgraphs of a graph $G = (V, E)$ given a partition $\Pi : [n] \mapsto [k]$ need to be built.
    To circumvent the costly download and upload between the CPU and GPU, the entire process is executed on the GPU\@.
    The algorithm works in $k$ iterations, where each iteration builds one subgraph $G'$.
    Algorithm~\ref{alg:build-subgraph} shows the pseudocode of one loop iteration.
    We define
    \[
        \mathbf{1}[P] =
        \begin{cases}
            1 & \text{if } P \text{ is true}\\
            0 & \text{otherwise}
        \end{cases}
    \]
    as an indicator function, which returns $1$ if the predicate $P$ is true and $0$ else.

    Each iteration consists of three phases.
    The first phase determines the number of vertices $n'$, edges $m'$ and the subgraph's weight $w'$ using three \emph{parallel\_reduce} loops.
    A vertex $v$ belongs to the subgraph with id $k'$ if $\Pi(v) = k'$ and likewise an edge $\{u, v\}$ belongs to it if $\Pi(u) = \Pi(v) = k'$.
    Lines 1 -- 3 show this phase.
    The second phase remaps the vertices~$v \in V_i$ from $[n]$ to $[n']$.
    This remapping is a technical requirement, since graph partitioners expect a continuous range of vertices.
    The mapping $M: [n] \mapsto [n']$ is also later needed to propagate the resulting partition to the original vertices.
    Lines $5 - 8$ show the pseudocode.
    Note that in Kokkos, a single \emph{parallel\_scan} is sufficient to construct the mapping.
    In the third phase, the subgraph is constructed.
    Lines 10 to 26 show the process.
    At first the new degree of each vertex is determined in parallel (Lines 10--13).
    Based on the new degrees the row pointer array of the graph can be determined using a prefix sum (Line 12).
    This assigns each vertex in the subgraph a continuous range, to place its edges.
    The variables $E'$ in Lines~15--17 are the corresponding arrays of the subgraph.
    At last the actual insertion of the edges into the subgraph is done in Lines 18--26.
    For each vertex $v$ that belongs to the subgraph, its neighborhood is iterated over.
    For each neighbor $u$ that also belongs to the subgraph, the edge is inserted into the arrays and the index $i$ is incremented.
    Note that this whole phase can also be implemented as one large \emph{parallel\_scan}.
    Subgraph creation usually takes less than 5\% of the total runtime of the hierarchical multisection algorithm.

    \begin{algorithm}[t]
        \caption{Build Subgraph}\label{alg:build-subgraph}
        \begin{algorithmic}[1]
            \Require Graph $G$, Partition $\Pi$, Subgraph ID $k'$
            \State $n' \gets \sum_{v \in V} \textbf{1}[\Pi(v) = k']$ \Comment{Phase 1}
            \State $w' \gets \sum_{v \in V} \textbf{1}[\Pi(v) = k']\, c(v)$
            \State $m' \gets \sum_{\{u,v\} \in \mathbb{E}} \textbf{1}[\Pi(u)=k' \land \Pi(v)=k']$
            \State $ $
            \State $B \gets \textsc{Zeros}(n)$ \Comment{Phase 2}
            \ParallelFor{$v \in V$}
            \State $B[v] \gets \textbf{1}[\Pi(v) = k']$
            \EndParallelFor
            \State $M \gets \textsc{PrefixSum}(B)$
            \State $ $
            \State $D \gets \textsc{Zeros}(n')$ \Comment{Phase 3}
            \ParallelFor{$v \in V$}
            \If{$\Pi(v) = k'$}
                \State $D[M(v)] \gets \sum_{u \in N(v)} \textbf{1}[\Pi(u)=k']$
            \EndIf
            \EndParallelFor
            \State $R' \gets \textsc{PrefixSum}(D)$ \Comment{Graph arrays}
            \State $E'_v \gets \textsc{Array}(m')$
            \State $E'_w \gets \textsc{Array}(m')$
            \State $E'_u \gets \textsc{Array}(m')$
            \ParallelFor{$v \in V$}
            \If{$\Pi(v) = k'$}
                \State $i \gets R'[M(v)]$
                \For{$u \in N(v)$}
                    \If{$\Pi(u) = k'$}
                        \State $E'_v[i] \gets M(u)$
                        \State $E'_w[i] \gets \omega(u, v)$
                        \State $E'_u[i] \gets M(v)$
                        \State $i \gets i + 1$
                    \EndIf
                \EndFor
            \EndIf
            \EndParallelFor
            \State \textbf{return} $G' = (R', E'_v, E'_w, E'_u),\ M$
        \end{algorithmic}
    \end{algorithm}

    The complete hierarchical multisection algorithm can be seen in Algorithm~\ref{alg:rec-hierarchical-multisection}.
    Here, it is implemented as a recursive function but an iterative approach would also be possible.
    The function takes as input a (sub)graph $G'$, the hierarchy $H$ and the current hierarchy level~$i$.
    Lines 2 and 3 handle the partitioning of the graph.
    At first the adaptive imbalance~$\epsilon'$ is computed using Equation~\ref{eq:adaptive-imbalance} and afterwards our reimplementation of \algo{Jet}~\cite{Gilbert24} is used to partition the graph on the GPU\@.
    It determines a partition $\Pi'$ into $a_i$ blocks with the specified $\epsilon'$ imbalance.
    If this was the last partitioning of this recursive call ($i = 1$) then the determined partition $\Pi'$ needs to be propagated to the actual partition $\Pi$ of $G$.
    Since this is only a technical process we omit its details and only outline the brief idea.
    For each created graph $G'$ keep a mapping $M'$ from $V'$ to the original vertices $V$ in $G$.
    Then for each $v \in V'$ set $\Pi(M'(v)) = \Pi'(v)$ into the final partition.
    The mapping $M'$ can be computed by utilizing the mapping $M$ from Algorithm~\ref{alg:build-subgraph}.
    If the subgraph $G'$ resides on a higher \hbox{level~($i > 1$)}, then the resulting subgraphs need to be constructed and further partitioned.
    In Line 7 the subgraphs are built using Algorithm~\ref{alg:build-subgraph}.
    Each subgraph is then recursively passed to the function, but with $i - 1$ it is indicated that the subgraphs reside one level further down in the hierarchy.
    Finally, the complete algorithm is invoked in Line 10 with the task graph $G_{\mathcal{C}}$, the hierarchy $H$ and the highest hierarchy level $\ell$.

    \begin{algorithm}[t]
        \caption{Recursive Hierarchical Multisection}\label{alg:rec-hierarchical-multisection}
        \begin{algorithmic}[1]
            \Require Graph $G$, Hierarchy $H=a_1:\ldots:a_{\ell}$, Imbalance $\epsilon$
            \Function{HM}{$G', H, i$}
                \State $\epsilon' \gets \textsc{AdapImb}(\ldots)$ \Comment{uses Equation~\ref{eq:adaptive-imbalance}}
                \State $\Pi' \gets \textsc{GPU-GP}(G', a_i, \epsilon')$        \Comment{GPU Graph Partitioner}

                \If{$i = 1$}
                    \State propagate $\Pi'$ to final partition $\Pi$
                    \State \textbf{return}
                \EndIf

                \For{$G'' \in \textsc{CreateSubgraphs}(G', \Pi')$}
                    \Comment{uses Algorithm~\ref{alg:build-subgraph}}
                    \State ${\textsc{HM}(G'', H, i - 1)}$
                \EndFor
            \EndFunction
            \State $\textsc{HM}(G_{\mathcal{C}}, H, \ell)$
        \end{algorithmic}
    \end{algorithm}

    \subsection{Integrated Mapping}\label{subsec:integrated-mapping}
    The second main contribution of this work is a GPU-based integrated mapping algorithm that incorporates the mapping directly into the graph partitioning.
    We follow the multilevel approach, but instead of minimizing the edge-cut, we minimize $J(\mathcal{C}, \mathcal{D}, \Pi)$.
    Remember that the multilevel approach first coarsens the graph into a sequence of progressively smaller graphs.
    The smallest graph is then partitioned and the coarsening is reverted.
    Refinement algorithms are used during uncoarsening to further optimize the objective.
    We will detail our methodology for each phase in the next few paragraphs.
    We closely follow the matching, coarsening, refinement and rebalancing methods used by the GPU-based graph partitioner \algo{Jet}~\cite{Gilbert24} and adapt them to process mapping.

    \textbf{Matching.}
    To determine a matching, the \emph{two-hop matching} presented by LaSalle~et~al.~\cite{LaSalle15} is adapted with a few minor changes.
    It starts with a heavy-edge matching algorithm.
    Instead of using $\omega(\{u, v\})$ as the rating of the edge, the rating function $\exp^{*2}(\{u, v\}) = \frac{\omega(\{u, v\})^2}{c(u)c(v)}$ presented by Holtgrewe~et~al.~\cite{Holtgrewe10} is used.
    Their experiments show that it leads to improved running times and solution quality in the context of graph partitioning.
    In parallel, each thread processes one vertex $v$.
    All unmatched neighbors are scanned and the vertex $u$ with the highest rating is saved as the preferred neighbor $p(v) = u$.
    To prevent identical ratings, a small amount of deterministic noise $\eta(\{u, v\})$ is added to each rating.
    It breaks any equalities, but is small enough to not influence the result.
    Vertices $u$ with $c(v) + c(u) > L_{\max}$ are excluded to prevent an imbalanced initial solution.
    Afterwards, each vertex $v$ is again processed in parallel.
    If $p(p(v)) = v$, i.e., vertex $v$ prefers $p(v)$ and $p(v)$ prefers $v$, then $v$ and $p(v)$ are matched to each other.
    The matching procedure is repeated for multiple rounds until no more matches are determined.
    If more than $75\%$ of vertices are matched then the matching stops.
    Else, leaf, twin and relative matchings are applied to increase the number of matched vertices.
    We adopt the same hashing and matchmaker vertex strategy of \algo{Jet} to improve performance.
    In process mapping, matching-based coarsening approaches are usually advantageous, since the task graphs stem from scientific simulations and have a mesh-like structure.
    Another approach would be to cluster the graph, however this is usually superior in unstructured workloads like social networks.

    \textbf{Contraction.}
    Edge contraction, to create the coarse graph, is based on a fine-grained per-vertex hashing scheme from \algo{Jet}~\cite{Gilbert24}, however, it is extended by using the extended CSR format.
    The overall idea is to assign each coarser vertex a hash array and the threads then hash the coarser edges into the respective array.
    The pseudocode can be seen in Algorithm~\ref{alg:gpu-coarsening}.
    Let $G=(V, E)$ be the current graph, $n_c$ the number of vertices in the coarse graph and $M : V \mapsto [n_c]$ map each vertex to its coarse counterpart.
    The algorithm first computes an upper bound on the neighborhood size of each coarse vertex by summing the degrees of all vertices assigned to it (Lines $1 - 3$).
    In Line 4 and 5 two hash arrays, $H_v$ and $H_w$, are allocated that will hold the edges and their weights.
    The offset array $O$ denotes the index range $[O[v], O[v + 1])$ where the edges of the coarse vertex $v$ are placed.
    Then, all edges $(u, v, w) \in \mathbb{E}$ are processed in parallel.
    If both endpoints contract to the same vertex, the edge becomes a self-loop and is discarded (Line 8).
    Otherwise, we need to insert $(M(v), w)$ into the adjacency of $M(u)$ in the interval $\mathcal{I}$.
    Note that adding $(M(u), w)$ into the adjacency of $M(v)$ is handled by the work unit with reverse edge $(v, u, w) \in \mathbb{E}$.
    Insertion is done in Line 10, which we will explain in the next paragraph.
    After the parallel loop ends, it remains to extract the actual graph from the arrays $H_v$ and $H_w$.
    This boils down to determining the true neighborhood sizes for all vertices, building the offset and extracting the edges by skipping all \textsc{Null} entries in $H_v$.
    This can be accomplished by two \emph{parallel\_scans}.

    \begin{algorithm}[t]
        \caption{GPU-based Contraction}\label{alg:gpu-coarsening}
        \begin{algorithmic}[1]
            \Require Graph $G$, Mapping $M$, \# Coarse Vertices $n_c$
            \State ${B \gets \textsc{Zero}(n_c)}$      \Comment{Upper Bounds}
            \ParallelFor{$v \in V$}
            \State ${B[M(v)] \gets B[M(v)] + |N(v)|}$
            \EndParallelFor
            \State ${H_v \gets \textsc{Null}(|E|)}$   \Comment{Vertices}
            \State ${H_w \gets \textsc{Zero}(|E|)}$    \Comment{Weights}
            \State ${O \gets \textsc{PrefixSum}(B)}$   \Comment{Offsets}
            \ParallelFor{$(u, v, w) \in \mathbb{E}$}
            \State \textbf{if} $M(u) = M(v)$ \textbf{then} \textsc{skip} \Comment{Self-loop discarded}
            \State ${\mathcal{I} \gets [O[M(u)], O[M(u) + 1])}$
            \State ${\textsc{InsertOrLookup}(H_v, H_w, \mathcal{I}, M(v))}$
            \EndParallelFor
            \State \textbf{return} ${\textsc{ExtractCSR}(B, H_v, H_w)}$
        \end{algorithmic}
    \end{algorithm}

    The most important part of the contraction is the \textsc{InsertOrLookup} function in Line 10.
    Each thread must place its coarse edge $(M(v), w)$ into $H_v$ and $H_w$ in the interval $\mathcal{I}$.
    In a serial algorithm one would first check if the edge $(M(v), w)$ already has an entry in $H_v$ and if so, add the weight to it.
    If there is no such entry then one would place the edge at a \textsc{Null} entry.
    However, in a parallel setting multiple threads add edges simultaneously and it must be ensured that edges are not overwritten and that duplicate edges are summed properly.
    To handle concurrent insertions, atomic compare-and-swap (CAS) operations can be utilized.
    A CAS operation $\text{CAS}(\textsc{ptr}, \textsc{expected}, \textsc{desired})$ compares the value at memory address~\textsc{ptr}, with the \textsc{expected} vale and only if equal, writes the \textsc{desired} value in its place.
    Regardless of whether the thread wrote \textsc{desired} or not, the previous value stored at \textsc{ptr} is always returned.
    The important part of the CAS operation is that the check and the write are executed as one atomic operation.
    Two threads can never read the same \textsc{expected} value and write different \textsc{desired} values.

    A thread now inserts its edge by executing a CAS operation on memory address $H_v[j]$ with the expected value of \textsc{Null} and the desired value $M(v)$.
    If $M(v)$ is returned, then the edge was already written in $H_v[j]$ and the weight $w$ is atomically added to $H_w[j]$.
    If \textsc{Null} is returned, then this thread just wrote $M(v)$ into $H_v[j]$ and $w$ is also atomically added to $H_w[j]$.
    If neither is returned, then $H_v[j]$ already holds a different edge.
    The thread will continue cycling through the interval~$\mathcal{I}$ until a valid position has been found.
    Since the neighborhood size of $M(u)$ was overestimated, at least one open slot will be available and the thread is able to place its edge.
    To reduce atomic contention on $H_v[j]$, each thread determines $j \in \mathcal{I}$ by hashing $M(v)$ and then cycling through the interval if necessary.

    \textbf{Initial Partitioning.}
    Initial partitioning is performed once the last coarsened graph has fewer than $8k$ vertices.
    We apply the hierarchical multisection~\cite{Kirchbach20, Schulz25} approach, explained in Section~\ref{subsec:hierarchical-multisection}, to acquire an initial mapping, but execute it entirely on the CPU\@.
    At this stage, the graph is sufficiently small that GPU execution offers no runtime advantage.
    Moreover, even a GPU-based graph partitioner like \algo{Jet} would internally quickly fall back to a CPU-based partitioner when presented with such small graphs.
    Also currently no GPU-based initial partitioning algorithms exist for HPMP or GP\@.
    To execute hierarchical multisection, we utilize a simple $k$-way graph partitioner.

    \textbf{Uncontraction.}
    Each coarser graph $G, G_1, \cdots, G_t$ is separately stored in memory so uncontracting the graph from $G_{i+1}$ to $G_i$ is achieved by simply deleting $G_{i+1}$.
    The mapping $\Pi$, however, has to be updated.
    Let $M$ be the mapping that was used to create the coarser graph $G_{i+1}$ from $G_i$.
    Then set in parallel $\Pi(v) = \Pi(M(v))$ for each $v \in G_i$.

    \textbf{Refinement.}
    We adapt the unconstrained label propagation of \algo{Jet}~\cite{Gilbert24} to use $J(\mathcal{C}, \mathcal{D}, \Pi)$ as the objective function instead of edge-cut.
    Classic label propagation visits all vertices in a random order and if moving the current vertex to a neighboring block improves the objective and adheres to the balance constraint, then the move is made immediately.
    Therefore, the objective value never decreases and the balance constraint is always fulfilled.
    In a parallel setting, one would like to determine for each vertex in parallel whether moving it is beneficial.
    This part can easily be parallelized.
    Then, the beneficial moves would ideally also be executed in parallel.
    However, when moving the vertices in parallel, difficulties arise.
    Each individual move might adhere to the balance constraint, but moving multiple vertices at once can quickly overload a block.
    Also, although each individual move might be beneficial, moving adjacent vertices in parallel can have an overall negative effect on the solution quality.
    This makes it inherently difficult to parallelize label propagation.

    \textsc{Jet}'s approach works in two phases.
    At first an unconstrained label propagation algorithm is used that determines beneficial vertex moves.
    In this phase the balance constraint is completely ignored and subsequently a second rebalancing phase is required.
    This phase aims to restore the balance, while keeping the loss in the objective function small.

    The unconstrained label propagation works in two phases.
    At first, for each vertex the best move to a neighboring block is determined.
    These moves are filtered and only non-negative moves are considered further.
    Each move is then reevaluated under a future approximate partition state.
    If the move stays positive, then it is finally saved to be executed.

    The pseudocode of our adapted refinement algorithm is provided in Algorithm~\ref{alg:label-propagation}.
    The refinement algorithm begins by determining the best target block $\Pi'(v)$ for each vertex $v$ by maximizing the gain $G_b(v)$ (see Equation~\ref{eq:gain}) across all neighboring blocks (Lines 4 -- 6).
    If indeed a move with non-negative gain has been found, then we insert the vertex into the list~$\mathcal{X}$ via an atomically incremented index (Line 7).

    \begin{algorithm}
        \caption{Unconstrained Label Propagation}\label{alg:label-propagation}
        \begin{algorithmic}[1]
            \Require Graph ${G}$, Mapping $\Pi$
            \State ${\Pi' \gets \Pi}$
            \State ${\mathcal{X} \gets \textsc{EmptyList}()}$ \Comment{list for first filter}
            \State ${i \gets 0}$ \Comment{atomic index}
            \ParallelFor{$v \in V$}
            \State ${B \gets \{\Pi(u) \mid u \in N(v) \}}$
            \State ${\Pi'(v) \gets \text{argmax}_{b \in B} G_{b}(v)}$
            \State \textbf{if} ${G_{\Pi'(v)}(v) \geq 0}$ \textbf{then} ${\mathcal{X}[i++] = v}$
            \EndParallelFor
            \State ${\mathcal{M} \gets \textsc{EmptyList}()}$ \Comment{list for actual moves}
            \State ${i \gets 0}$
            \ParallelFor{$v \in \mathcal{X}$}
            \State \textbf{if} ${\mathbb{G}_{\Pi'(v)}(v) \geq 0}$ \textbf{then} ${\mathcal{M}[i++] = v}$
            \EndParallelFor
            \State \textbf{return} ${\mathcal{M}, \Pi'}$
        \end{algorithmic}
    \end{algorithm}

    Originally, \algo{Jet} applies a more relaxed filter that also retains negative vertex moves.
    A vertex $v$ must only satisfy \hbox{$G_b(v) \geq 0 \lor -G_b(v) < \lfloor c \cdot \operatorname{conn}(v, \Pi(v)) \rfloor$}.
    The value $\operatorname{conn}(v, \Pi(v))$ denotes the sum of edge weights connecting $v$ and block $\Pi(v)$.
    Negative vertex moves are allowed to pass this filter, such that the algorithm may escape local minima.
    The criterion is suitable when minimizing the edge-cut, as both $G_b(v)$ and $\operatorname{conn}(v, \Pi(v))$ depend solely on edge weights and are thus directly comparable.
    In contrast, for process mapping, $G_b(v)$ contains the additional communication factors and is therefore often much greater than $\operatorname{conn}(v, \Pi(v))$.
    This renders the original filter ineffective.
    Therefore only non-negative moves~$G_b(v) \geq 0$ are allowed to pass the filter and be inserted into $\mathcal{X}$, as it performed best in our preliminary experiments.
    A more sophisticated filtering criterion may further improve solution quality and remains a topic for future work.

    The list $\mathcal{X}$ now holds all non-negative vertex moves.
    In a serial algorithm, one could now sort the list and execute the best vertex moves first.
    Moves further to the end of the list could then be reevaluated under the new partition state and only be executed if they are positive.
    However, in a parallel setting this is not straightforward.
    Firstly, sorting on a GPU is expensive and should be avoided and secondly, executing one move after the other is inherently serial.
    Gilbert~et~al.~\cite{Gilbert24} therefore first only implicitly sort the list $\mathcal{X}$ and then each vertex move is re-evaluated under an approximate future partition state in parallel.
    The list $\mathcal{X}$ is unordered, but we can define an ordering on its elements via
    \[
        \operatorname{ord}(u) \begin{cases}
                                  < \operatorname{ord}(v) & \text{if } u \in \mathcal{X} \land G_{\Pi'(u)}(u) > G_{\Pi'(v)}(v), \\
                                  < \operatorname{ord}(v) & \text{if } u \in \mathcal{X} \land G_{\Pi'(u)}(u) = G_{\Pi'(v)}(v) \land u < v, \\
                                  > \operatorname{ord}(v) & \text{otherwise}.
        \end{cases}
    \]
    Vertices with a high gain are at the front of the list, while vertices with lower gain are towards the end of the list.
    By querying ${G_{\Pi'(v)}(v) \geq 0}$ we can determine whether a vertex was inserted into $\mathcal{X}$.
    It is now assumed that vertices with a lower ordering would have moved in a serial algorithm before a vertex with a higher ordering.
    All vertex moves in $\mathcal{X}$ are now reevaluated in parallel under an approximate future mapping state.
    We denote the reevaluated gain by $\mathbb{G}_{\Pi'(v)}(v)$ and it is determined just like the gain $G_{\Pi'(v)}(v)$ (Equation~\ref{eq:gain}), only the assigned block of some neighbors change.
    If a vertex $v$ has a neighbor $u$ with $\operatorname{ord}(u) < \operatorname{ord}(v)$ and vertex $u$ is scheduled to move to block $\Pi'(u)$, then $u$ is assumed to have already moved when recomputing the gain~$\mathbb{G}_{\Pi'(v)}(v)$.
    Each vertex $v \in \mathcal{X}$ that has a non-negative gain $\mathbb{G}_{\Pi'(v)}(v)$ is then added to the final list of moves (Lines 9 -- 11).
    To prevent oscillation of vertices, all vertices in the final list $\mathcal{M}$ are locked for the next iteration of label propagation.
    The results are also cached and if the neighborhood of a vertex did not change, then its result is reused.
    Both of these techniques are not shown in Algorithm~\ref{alg:label-propagation} for visual clarity.
    The function returns the list $\mathcal{M}$ and the destination mapping $\Pi'$.

    \textbf{Rebalancing.}
    The rebalancing is also adapted from \algo{Jet}~\cite{Gilbert24}, which splits its rebalancing into a weak and strong part.
    The idea is to determine for each vertex in an overloaded block the move with minimal loss to an underloaded block.
    All moves are approximately sorted and then as few vertices as necessary are moved.
    Weak rebalancing aims to keep the loss small, but might need multiple iterations, while strong rebalancing balances in one iteration, but has a much greater loss.

    The \emph{weak} rebalancing starts by determining for each vertex $v$ in an overloaded block the move with maximum gain (or equally minimum loss) $G_b(v)$ into a neighboring block~$B$ that has a weight less than the threshold $\sigma \leq L_{\max}$.
    The threshold $\sigma$ is chosen such that there is a small dead zone between $\sigma$ and $L_{\max}$, so $B$ can take multiple vertices without instantly being overloaded.
    In our implementation we set $\sigma = L_{\max} - 100$, just like \algo{Jet}.
    If for a vertex no such neighboring block exists, a random block that satisfies the threshold is chosen.
    Heavy vertices $v$ that weigh more than $1.5(c(\Pi(v)) - c(V) / k)$ are not allowed to move during rebalancing.

    Ideally, each block would store all its outgoing vertex moves in decreasing order in a list and then a prefix of the list could be chosen, such that the block becomes balanced.
    This would grant a relatively small loss based on the current partition.
    However, sorting on a GPU is expensive, so instead an approximation of the sorted list is utilized.
    Each negative vertex move is assigned a slot in a $\log_2$--spaced bucket list.
    A vertex $v$ with negative gain $G_b(v)$ would therefore be stored in bucket $i$ with $i \leq \log_2(-G_b(v)) < i+1$.
    This gives an increased resolution for moves with a comparatively small loss, which minimizes the loss when only a small number of vertices need to be moved.
    Strictly positive vertex moves are assigned to a special $+$-bucket and 0-gain moves are placed in a special 0-bucket.

    Instead of first building the approximation list explicitly and then determining the prefix of vertex moves, one can also use a per-vertex decision process.
    For each vertex, when it is added to a bucket it will store the accumulated weight of vertices already inside of its bucket.
    After all vertices have been processed, a prefix sum over all bucket weights is computed.
    Through this, each vertex can now determine the sum of the vertex weights in earlier buckets and it knows how much vertex weight will be moved in its own bucket, before the vertex itself is moved.
    If this weight is smaller than $c(B) - L_{\max}$, then this vertex needs to be moved.

    \begin{algorithm}[t]
        \caption{Weak Rebalancing}\label{alg:weak-rabalancing}
        \begin{algorithmic}[1]
            \Require Graph $G=(V, E)$, Mapping $\Pi$
            \State ${\Pi' \gets \Pi}$
            \ParallelFor{$v \in V$}
            \State \textbf{if} $c(\Pi(v)) \leq L_{\max}$ or $v$ too heavy \textbf{do} \textsc{Skip}
            \State ${B \gets \{\Pi(u) \mid u \in N(v) \land c(\Pi(u)) \leq \sigma \}}$
            \If{$B = \emptyset $}
                \State $B \gets \{ \text{rnd Block b with }c(b) \leq \sigma \}$
            \EndIf
            \State $g \gets \max_{b \in B} G_{b}(v)$
            \State ${\Pi'(v) \gets \text{argmax}_{b \in B} G_{b}(v)}$
            \EndParallelFor
            \State ${\mathcal{M} \gets \textsc{EmptyList}()}$ \Comment{list for actual moves}
            \State ${i \gets 0}$ \Comment{atomic index}
            \ParallelFor{$v \in V$}
            \State \textbf{if} ${\mathbb{G}_{\Pi'(v)}(v) \geq 0}$ \textbf{then} ${\mathcal{M}[i++] = v}$
            \EndParallelFor
            \State \textbf{return} ${\mathcal{M}, \Pi'}$
        \end{algorithmic}
    \end{algorithm}

    By approximately sorting the lists by gain, and only moving as few vertices as possible, the overall combined loss stays small.
    However, there is no safeguard against the destination blocks becoming overloaded.
    While the threshold $\sigma$ helps in preventing this, it does not guarantee it.
    Therefore, the destination blocks might become overloaded and another iteration is required.
    The \emph{strong} rebalancing operates in a very similar manner to weak rebalancing but takes extra precaution to not overload the destination blocks.
    Vertices that would overload a destination block are reassigned to other underloaded blocks.
    However, the vertices are not necessarily connected to their new destination block, which results in a much greater loss.

    We adapted the method to use $J(\mathcal{C}, \mathcal{D}, \Pi)$ as the objective function.
    Although this increases computational complexity, we observed no improvement in solution quality.
    Therefore, we use the standard rebalancing that minimizes edge-cut, which performs equally well but is faster to compute.
    One explanation is that edge-cut loss and communication cost loss correlate, so moves with small edge-cut loss also incur small communication cost loss.
    So regardless of the objective function, similar sets of vertices are moved.
    Additionally, any poor rebalancing move is likely corrected during the label propagation phase due to its high reversal gain.

    \begin{algorithm}[t]
        \caption{Overall Refinement Algorithm}\label{alg:overall-refinement}
        \begin{algorithmic}[1]
            \Require Graph $G=(V, E)$, Mapping $\Pi$
            \State $\Pi' \gets \Pi$
            \State $i \gets 0, \quad i_w \gets 0$
            \While{$i < 12$}
                \State $i \gets i + 1$
                \If{$\operatorname{maxImb}(G, \Pi') \leq L_{\max}$}
                    \State $\mathcal{M},$ $\Pi'' \gets \operatorname{UnconLabelProp}(G, \Pi')$
                    \State $i_w \gets 0$
                \Else
                    \If{$i_w < 2$}
                        \State $\mathcal{M}, \Pi'' \gets \operatorname{weakReb}(G, \Pi')$
                        \State $i_w \gets i_w + 1$
                    \Else
                        \State $\mathcal{M}, \Pi'' \gets \operatorname{strongReb}(G, \Pi')$
                        \State $i_w \gets 0$
                    \EndIf
                \EndIf
                \State $\Pi' \gets \operatorname{Move}(\mathcal{M}, \Pi'')$
                \If{$\operatorname{maxImb}(G, \Pi') \leq L_{\max}$}
                    \State \textbf{if} $J(\mathcal{C}, \mathcal{D}, \Pi') < J(\mathcal{C}, \mathcal{D}, \Pi)$ \textbf{then} $\Pi \gets \Pi'$
                    \State \textbf{if} $J(\mathcal{C}, \mathcal{D}, \Pi') < \phi J(\mathcal{C}, \mathcal{D}, \Pi)$ \textbf{then} $i \gets 0$
                \Else
                    \If{$\operatorname{maxImb}(G, \Pi') < \operatorname{maxImb}(G, \Pi)$}
                        \State $\Pi \gets \Pi', \quad i \gets 0$
                    \EndIf
                \EndIf
            \EndWhile
        \end{algorithmic}
    \end{algorithm}

    \textbf{Overall Refinement Algorithm.}
    The complete refinement procedure is summarized in Algorithm~\ref{alg:overall-refinement}.
    The algorithm iteratively alternates between unconstrained label propagation and rebalancing (depending on the current imbalance) until at least 12 iterations have been executed.
    This guarantees that at least three full cycles of label propagation, then two weak rebalancing steps and at last one strong rebalancing can be executed.
    A copy $\Pi'$ is used and the actual mapping $\Pi$ is only updated if an improved mapping has been found.
    In each iteration, if the current partition $\Pi'$ satisfies the imbalance constraint (Line 5), the algorithm performs unconstrained label propagation to reduce the communication cost (Line 6).
    Otherwise, the algorithm invokes weak or strong rebalancing to improve the balance (Lines 8 -- 15).
    After each iteration, the partition $\Pi'$ is updated (Line 16), and the best feasible solution encountered so far is saved.
    The process terminates when no significant improvement in either objective quality or balance has been achieved within the iteration limits.
    The hyperparameter $\phi = 0.999$ specifies the minimum relative improvement of $J(\mathcal{C}, \mathcal{D}, \Pi)$ that has to be achieved in each iteration.
    The constants 12, 2 and $\phi$ are transferred from \algo{Jet}.

    During the algorithm, an additional structure (not shown in the pseudocode) stores for each vertex $v$ all neighboring blocks and the sum of edge weights to those blocks.
    This structure is used to efficiently compute gains during label propagation and rebalancing.
    For each vertex it holds a hash array of size $\min(|N(v)|, k)$ to store all possible connected blocks.
    In contrast to \algo{Jet}, the structure is initialized by using the extended CSR format.
    By using an edge-parallel loop each thread cycles through the vertex hash array and uses \textsc{CAS} operations to find a valid entry.
    When moving the vertices (Line 15) the structure must be updated.
    This is done by either completely resetting the affected vertices' hash arrays and filling them from scratch, or by first removing old connections and then adding new connections.

    \section{Experimental Evaluation}\label{sec:experiments}
    All of our algorithms\footnote{\href{https://github.com/HenningWoydt/GPU-HeiPa}{\faGithub\ / GPU-HeiPa}} are implemented in C++ 20 and utilize the Kokkos library~\cite{Edwards14} version 5.0.0 for the GPU code.
    It compiles the host-side code with GCC 11.4.0 and the GPU code with NVIDIA Cuda compiler 12.6.77.
    We compile with the highest optimization (\textsc{-O3}).

    \textbf{System.}
    The experiments are conducted on a machine that has an Intel\textsuperscript{\textregistered} Xeon\textsuperscript{\textregistered} w5-3435X that runs at a max frequency of $3.1$ GHz.
    The system features 128 GB of main memory.
    The GPU is an NVIDIA\textsuperscript{\textregistered} GeForce RTX\texttrademark{} 4090 with 24 GB of memory.
    It has 16,384 Cuda cores with a boost frequency of $2.5$ GHz.
    The system runs Ubuntu~22.04.4.

    \begin{table}[t]
    \caption{Benchmark instances with their corresponding numbers of vertices and edges.}
    \label{tab:benchmark-instances}
    \centering
    \setlength{\tabcolsep}{5pt}
    \renewcommand{\arraystretch}{0.95}
    \resizebox{\columnwidth}{!}{
        \begin{tabular}{@{}lrrlrrlrr@{}}
            \toprule
            \multicolumn{3}{c}{\textbf{SuiteSparse}} &
            \multicolumn{3}{c}{\textbf{Other}} &
            \multicolumn{3}{c}{\textbf{Walshaw}} \\
            \cmidrule(lr){1-3}
            \cmidrule(lr){4-6}
            \cmidrule(lr){7-9}
            Graph & $|V|$ & $|E|$ &
            Graph & $|V|$ & $|E|$ &
            Graph & $|V|$ & $|E|$ \\
            \midrule
            cop20k\_A      & 99\,843  & $\approx$1.3M & afshell9   & 504\,855 & $\approx$8.5M  & 598a      & 110\,971 & 741\,934 \\
            2cubes\_sphere & 101\,492 & 772\,886      & thermal2   & $\approx$1.2M & $\approx$3.7M & fe\_ocean & 143\,437 & 409\,593 \\
            thermomech\_TC & 102\,158 & 304\,700      & nlr        & $\approx$4.2M & $\approx$12.5M & 144       & 144\,649 & $\approx$1.1M \\
            cfd2           & 123\,440 & $\approx$1.5M & deu        & $\approx$4.4M & $\approx$5.5M  & wave      & 156\,317 & $\approx$1.1M \\
            boneS01        & 127\,224 & $\approx$3.3M & del23      & $\approx$8.4M & $\approx$25.2M & m14b      & 214\,765 & $\approx$1.7M \\
            Dubcova3       & 146\,689 & $\approx$1.7M & rgg23      & $\approx$8.4M & $\approx$63.5M & auto      & 448\,695 & $\approx$3.3M \\
            bmwcra\_1      & 148\,770 & $\approx$5.2M & del24      & $\approx$16.8M & $\approx$50.3M &           &          &             \\
            G2\_circuit    & 150\,102 & 288\,286      & rgg24      & $\approx$16.8M & $\approx$132.6M &           &          &             \\
            shipsec5       & 179\,860 & $\approx$5.0M & europe\_osm & $\approx$50.9M & $\approx$54.1M &           &          &             \\
            cont-300       & 180\,895 & 448\,799      &            &                &                &           &          &             \\
            \bottomrule
        \end{tabular}
    }
\end{table}

    \textbf{Benchmark Instances.}
    To make our work more comparable, we choose the same benchmark instances as in~\cite{Schulz25, Faraj20, Heuer23, Schulz17, Kirchbach20}.
    They can be seen in Table~\ref{tab:benchmark-instances}.
    The small instances come from the SuiteSparse Matrix Collection~\cite{Davis11}.
    These represent task graphs from real applications.
    Six graphs were taken from Walshaws's benchmark archive~\cite{Soper04}, which holds instances for benchmarking graph partitioning algorithms.
    The graphs \texttt{afshell9}, \texttt{thermal2}, \texttt{nlr} are from the matrix and numerical section of the 10\textsuperscript{th} DIMACS Implementation Challenge~\cite{Bader14}.
    The graphs \texttt{deu} and \texttt{europe\_osm} are large road networks of Germany and Europe~\cite{Delling09}.
    Graphs \texttt{rgg23} and \texttt{rgg24} are random geometric graphs with $2^{23}$ and $2^{24}$ vertices respectively.
    Each vertex represents a random point in the unit square, and if their distance is less than $0.55\sqrt{\ln n / n}$ they are connected via an edge.
    Graphs \texttt{del23} and \texttt{del24} also have $2^{23}$ and $2^{24}$ vertices, with each vertex representing a point in the unit square.
    The edges are connected such that the graph is a Delaunay triangulation.

    \textbf{Experimental Setup.}
    As the hierarchy, we choose \hbox{$H = 4 : 8 : \{1, \ldots, 6\}$} and as the distance, we choose \hbox{$D = 1 : 10 : 100$}.
    The imbalance parameter is set to 3\% (\hbox{$\epsilon = 0.03$}).
    Each algorithm is run five times with different seeds and the \emph{running time} and \emph{communication cost} $J(\mathcal{C}, \mathcal{D}, \Pi)$ are averaged across the seeds.
    We do not include the time to initially read the graph from disk and constructing it on the CPU, and also not the time to write the partition to disk.
    These steps are conceptually the same for all algorithms.

    \textbf{Performance Profiles.}
    Performance Profiles~\cite{Dolan02} enable us to analyze the quality of multiple algorithms across multiple instances.
    Let $\mathcal{A}$ be the set of all algorithms and let $\mathcal{I}$ be the set of all instances.
    The quality of algorithm $A \in \mathcal{A}$ on instance $I \in \mathcal{I}$ is $q_{A}(I)$ and $\text{Best}(I) = \min_{A \in \mathcal{A}}q_{A}(I)$ denotes the best quality across all algorithms on instance $I$.
    For each algorithm $A$, a performance profile shows the fraction of instances on the $y$-axis, for which $q_A(I) \leq \tau \text{Best}(I)$, where $\tau \geq 1$ is on the $x$-axis.
    For $\tau = 1$, the profile shows the fraction of instances that each algorithm solved with the best solution.
    Algorithms that perform well have a high fraction of instances solved with minimal $\tau$ values, they are towards the top-left corner.
    Sub-optimal algorithms will have small fractions at minimal $\tau$ values, and only achieve higher fractions with greater $\tau$ values.

    \subsection{Results}\label{subsec:results}
    First, we analyze our integrated mapping (\algo{GPU-IM}) and hierarchical multisection (\algo{GPU-HM}) algorithms.
    \algo{Jet}~\cite{Gilbert24}, which is utilized in \algo{GPU-HM}, also has an \algo{ultra} configuration, which executes the complete refinement algorithm 18 times.
    This yields better solution quality at the cost of runtime.
    We abbreviate our approach that uses the \textsc{ultra} variant as \algo{GPU-HM-ultra}.
    In the end, we compare our methods against the established algorithms \algo{SharedMap}~\cite{Schulz25} and \algo{IntMap}~\cite{Faraj20}, which are CPU-based counterparts to hierarchical multisection and integrated mapping.

    \begin{figure*}[t]
        \centering
        \begin{minipage}{0.49\textwidth}
            \centering
            \includegraphics[width=0.9\linewidth]{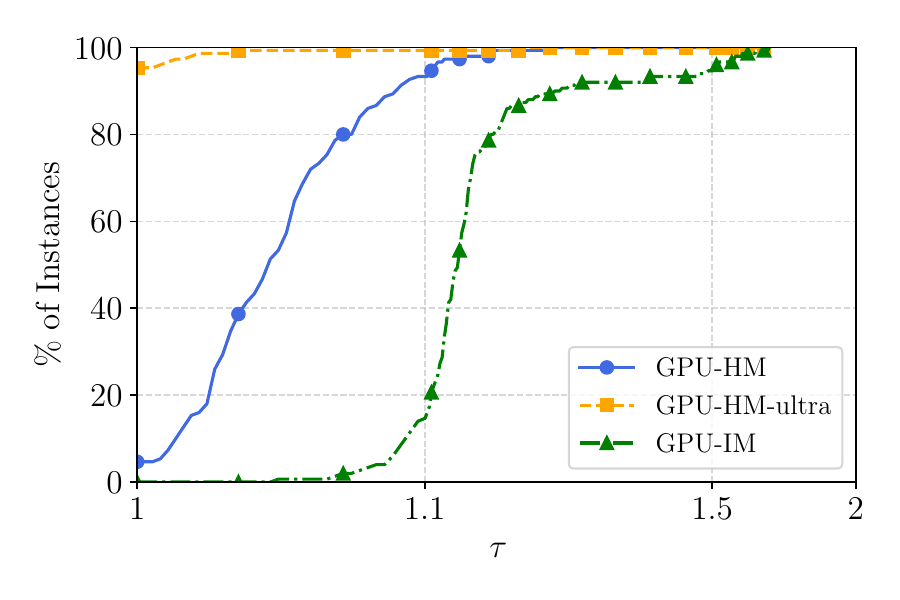}
        \end{minipage}
        \hfill
        \begin{minipage}{0.49\textwidth}
            \centering
            \includegraphics[width=0.9\linewidth]{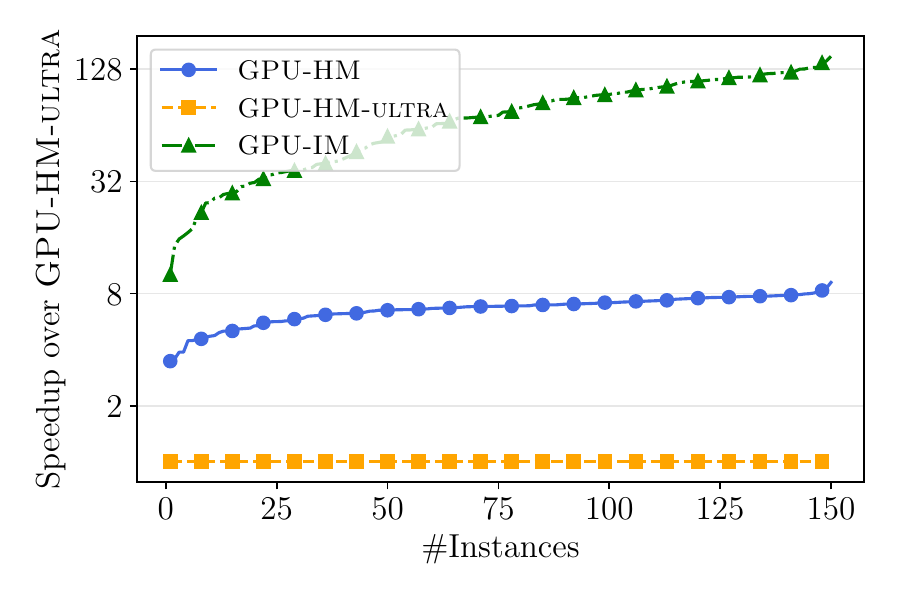}
        \end{minipage}

        \caption{Performance profile comparing the solution quality (left) and a speedup plot (right) of our proposed algorithms.}
        \label{fig:own}
    \end{figure*}

    \subsection{Own Comparison}\label{subsec:own-comparison}
    Figure~\ref{fig:own} shows the performance profile and the speedup of our algorithms across all instances.
    \algo{GPU-HM-ultra} serves as a baseline to measure the speedup as it achieves the lowest communication cost.
    The default \algo{GPU-HM} approach reaches a maximum speedup of $9.1\times$ with a geometric mean speedup of $6.5\times$.
    However, the speedup comes with an increase in communication cost.
    While the \algo{ultra} variant finds the best solution on $95.3\%$ of instances, \algo{GPU-HM} does so on only $4.7\%$.
    On average, the communication cost of \algo{GPU-HM-ultra} is just $0.2\%$ above the best solution, whereas \algo{GPU-HM} is $5.1\%$ worse.
    \textsc{GPU-IM} is the fastest of our algorithms with a maximum speedup of $150.1\times$ and a geometric mean speedup of $64.9\times$.
    However, it did not achieve the lowest communication cost on any instance and has, on average, $17.4\%$ higher communication cost than the best solution.

    Table~\ref{tab:phase-breakdown} reports the average percentage of runtime \textsc{GPU-IM} spends in its major phases.
    The hierarchical multisection approaches spend more than $95\%$ of runtime in the partitioning, so we do not analyze it any further.
    The results in the table are grouped by graph size: small graphs with fewer than one million vertices and large graphs with more than one million vertices.
    For smaller graphs, the most time consuming phase is the refinement phase where the algorithm almost spends two-thirds of the runtime.
    Both the coarsening and the initial partitioning phase take about $13\%$ of runtime, and the other phases have a negligible impact on the running time.
    For larger graphs the distribution is more balanced.
    Notably, refinement and rebalancing is again the most time consuming phase, with about $45\%$ of total runtime.
    The initial partitioning now also becomes negligible, however coarsening and contraction take about $11\%$ each.
    Interestingly all miscellaneous operations, which also include uploading and downloading the initial graph and the resulting partition from the CPU to the GPU and vice versa, is the second greatest runtime consumption.
    This phase is mostly hardware dependent and cannot be sped up by further algorithm improvements.
    For the largest and smallest graphs, namely \textsc{cop20k\_A} and \textsc{europe\_osm}, the absolute running time of each phase is shown in milliseconds for the hierarchy $4:8:6$.

    \begin{table}[t]
    \centering
    \caption{Runtime distribution across \algo{GPU-IM} algorithm phases. The last two columns hold the absolute running times of each phase in milliseconds.}
    \label{tab:phase-breakdown}
    \setlength{\tabcolsep}{6pt}
    \renewcommand{\arraystretch}{1.0}
    \begin{tabular}{@{}lrrrr@{}}
        \toprule
        \textbf{Phase} & \textbf{Small} & \textbf{Large} & \textbf{cop20k\_A} (ms) & \textbf{europe\_osm} (ms) \\
        \midrule
        Coarsening     & 13.02\% & 11.59\% & 4.351  & 41.020 \\
        Contraction    & 3.49\%  & 11.16\% & 1.010  & 38.694 \\
        Init.\ Part.   & 13.85\% & 4.23\%  & 11.116 & 7.244  \\
        Uncontr.       & 0.14\%  & 0.24\%  & 0.046  & 1.523  \\
        Refine + Reb.  & 65.22\% & 45.53\% & 24.359 & 116.598 \\
        Misc           & 4.28\%  & 27.24\% & 1.193  & 115.469 \\
        \midrule
        \textbf{Total} & \textbf{100\%} & \textbf{100\%} & \textbf{42.076} & \textbf{320.550} \\
        \bottomrule
    \end{tabular}
\end{table}

    \subsection{CPU-based Comparison}\label{subsec:cpu-based-comparison}
    We compare our algorithms to the CPU-based solvers \algo{SharedMap}~\cite{Schulz25} and \algo{IntMap}~\cite{Faraj20}, since they performed well in the experimental evaluation of~\cite{Schulz25}.
    \algo{SharedMap} employs the hierarchical multisection approach, just like \algo{GPU-HM}.
    It internally uses \algo{KaFFPa}~\cite{Sanders11} for its graph partitioning.
    \algo{IntMap}, in contrast, minimizes $J(\mathcal{C}, \mathcal{D}, \Pi)$ via the multilevel scheme, making it conceptually similar to \algo{GPU-IM}.
    Both algorithms have a \algo{Fast} and \algo{Strong} configuration which we denote by the suffix \algo{-F} and \algo{-S}, respectively.
    These configurations trade speed for solution quality.
    For comparison, we focus on \algo{GPU-HM-ultra}, which achieves the highest solution quality, and \algo{GPU-IM}, which provides the lowest running time.
    \algo{GPU-HM} is omitted for better visual clarity, as its performance lies between the two.

\begin{figure*}[t]
        \centering
        \begin{minipage}{0.49\textwidth}
            \centering
            \includegraphics[width=0.9\linewidth]{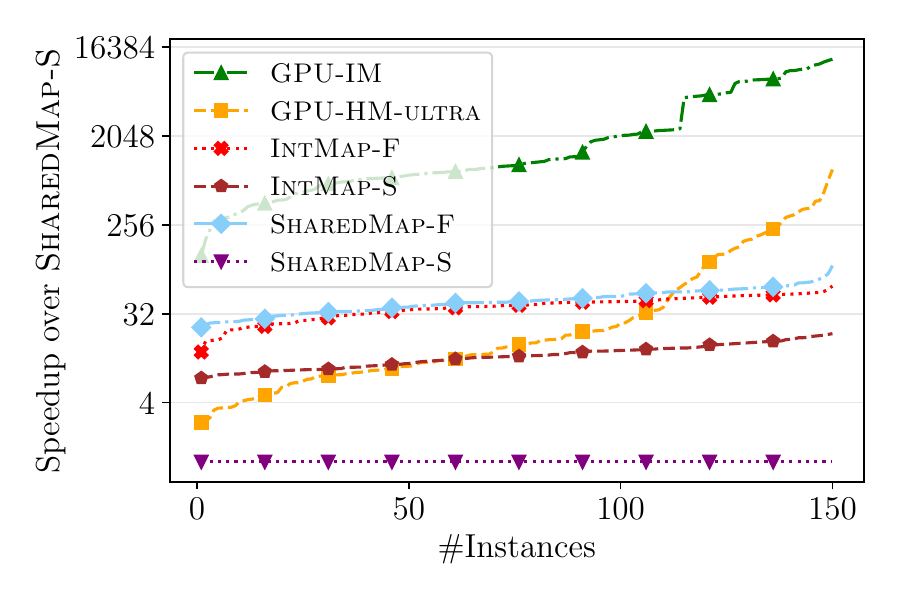}
        \end{minipage}
        \hfill
        \begin{minipage}{0.49\textwidth}
            \centering
            \includegraphics[width=0.9\linewidth]{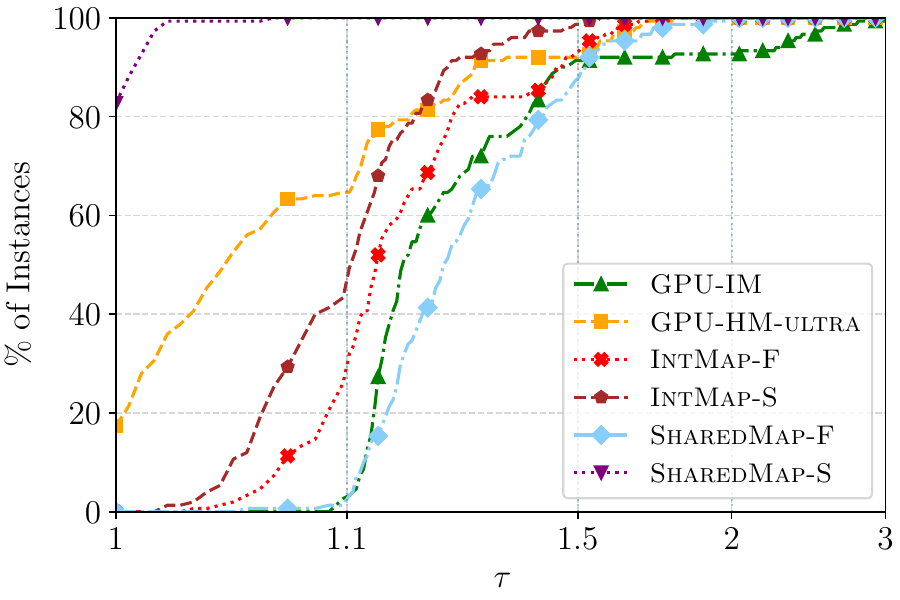}
        \end{minipage}

        \caption{Comparison of the speedup (left) and the performance profile of achieved communication costs (right) of our proposed algorithms \algo{GPU-HM-ultra} and \algo{GPU-IM} to CPU-based algorithms \algo{SharedMap}~\cite{Schulz25} and \algo{IntMap}~\cite{Faraj20}.}
        \label{fig:all}
    \end{figure*}

    Figure~\ref{fig:all} shows the speedup over \algo{SharedMap-S} and the performance profile.
    Among all algorithms, \algo{SharedMap-S} achieves the highest solution quality, obtaining the best solution on 124 instances ($82.7\%$). %
    However, this comes at the cost of runtime, as \algo{SharedMap-S} is also by far the slowest algorithm.
    \algo{GPU-HM-ultra} finds the best solution on the remaining 26 instances ($17.3\%$).
    All other algorithms are not able to determine a best solution on any instance.
    Ranked by average additional cost over the best solution, the algorithms are ordered as follows: \algo{SharedMap-S} ($0.2\%$), \algo{GPU-HM-ultra} ($12.2\%$), \algo{IntMap-S} ($14.4\%$), \algo{IntMap-F} ($20.9\%$), \algo{SharedMap-F} ($30.8\%$), and \algo{GPU-IM} ($33.1\%$).
    \algo{SharedMap-F} and \algo{GPU-IM} are at most $1.99\times$ and $3.07\times$ worse than the best-known solution, respectively.

    In terms of runtime, \algo{GPU-IM} is the best algorithm overall, achieving a maximum speedup of $12376.9\times$ and a geometric mean speedup of $1454.6\times$ compared to \algo{SharedMap-S}.
    In fact, it is the fastest algorithm on all 150 instances.
    While \algo{GPU-HM-ultra} achieves a maximum speedup of $934.7\times$, its geometric mean speedup is only $22.4\times$.
    The remaining mean speedups are $42.7\times$ for \algo{SharedMap-F}, $36.7\times$ for \algo{IntMap-F}, and $11.7\times$ for \algo{IntMap-S}.

    Our algorithms have two clear CPU-based counterparts.
    \algo{GPU-HM-ultra} outperforms \algo{IntMap-S} in terms of solution quality and speed.
    It finds better solutions on 117 out of 150 instances ($78.0\%$) and is also faster on 93 instances ($62.0\%$).
    When split into small and large graphs, with fewer and more than one million vertices respectively, \algo{GPU-HM-ultra} is only faster on 45 out of 102 small instances, but faster on all 48 large instances with a mean speedup of $12.6\times$.
    When inspecting the communication cost, \algo{GPU-HM-ultra} has better solution quality on 95 small instances, while only being better on 22 large instances.
    Similarly, \algo{GPU-IM}'s solution quality lies between \algo{SharedMap-F} and \algo{IntMap-F}.
    On smaller graphs \algo{GPU-IM} finds better solutions than \algo{SharedMap-F} on 70 instances, but only on 5 instances when compared to \algo{IntMap-F}.
    For larger graphs this ratio worsens as \algo{GPU-IM} only finds better solutions on 14 and 10 instances when compared to \algo{SharedMap-F} and \algo{IntMap-F} respectively.
    However, \algo{GPU-IM} is much faster than both algorithms with a mean speedup of $55.8\times$ and a max speedup of $221.6\times$ over \algo{SharedMap-F}.
    Compared to \algo{IntMap-F}, the mean speedup is $70.8\times$ and the max speedup is $243.1\times$.

    In summary, \algo{GPU-HM-ultra} outperforms \algo{IntMap-S} in both solution quality and running time.
    While being about $20$ times faster than \algo{SharedMap-S}, it cannot reach its state-of-the-art solution quality.
    \algo{GPU-IM} achieves a mean speedup of $1454\times$ compared to \algo{SharedMap-S}, but its communication cost is on average $33.1\%$ higher.
    Nonetheless, its quality is comparable to \algo{IntMap-F} and \algo{SharedMap-F} while being about $60$ times faster.

    \subsection{GPU-based Comparison}\label{subsec:gpu-based-comparison}
    While no other GPU-based algorithms currently exist for solving the GPMP, we can compare our algorithms to \algo{Jet}~\cite{Gilbert24}.
    Note that \algo{Jet} solves the graph partitioning problem, which corresponds to using the distance vector $D = 1 : \ldots : 1$.
    Since we test the distance $D = 1 : 10 : 100$, \algo{Jet}'s resulting partitions are unfit and the corresponding communication costs are worse than those of our algorithms.
    In terms of runtime, however, \algo{Jet} is comparable to \algo{GPU-IM} as they both follow the multilevel scheme.

    \textsc{Jet}'s determined partitions incur on average $45.3\%$ additional communication cost over those by \algo{GPU-IM}.
    Compared to \algo{SharedMap-S} this is an average increase of $90.3\%$.
    As explained it is expected that the produced partitions are unfit for process mapping, which highlights the need for a dedicated solver.
    Note that the quality of the \algo{ultra} configuration is even worse.
    This can be explained because decreasing the edge-cut does not necessarily decrease the communication cost.
    Trading two cut edges for one cut edge is always beneficial (assuming uniform edge weights) when optimizing the edge-cut, however this is not inherently true for process mapping.
    Trading two cut edges on neighboring PEs for one cut edge that connects two islands has an overall detrimental effect.
    In terms of runtime, \algo{GPU-IM} is slightly faster than \algo{Jet}.
    On small graphs, with fewer than one million vertices, the geometric mean speedup is $1.43\times$, with a minimum and maximum speedup of $0.21\times$ and $1.95\times$.
    For larger graphs, with more than one million vertices, the mean speedup increases to $1.56\times$.
    The minimum and maximum speedup are $1.21\times$ and $2.22\times$.
    We primarily attribute the small speed gains to the use of the extended CSR format and miscellaneous technical improvements.
    In summary, \algo{GPU-IM} has a geometric mean speedup of $1.47\times$ over \algo{Jet} and therefore delivers equal speed to current state-of-the-art GPU-based graph partitioners.

    \section{Conclusion}\label{sec:conclusion}
    Process mapping assigns vertices of a task graph to the processing elements (PEs) of a supercomputer to balance workload and minimize communication costs.
    Motivated by the success of GPU-based graph partitioners, we proposed two GPU-accelerated algorithms for this problem.
    The first, \algo{GPU-HM}, applies hierarchical multisection~\cite{Kirchbach20} using a reimplementation of the GPU-based graph partitioner \algo{Jet}~\cite{Gilbert24}.
    The second, \algo{GPU-IM}, integrates process mapping directly into a multilevel framework, adapting \algo{Jet}’s label propagation and rebalancing.
    \algo{GPU-HM-ultra} achieves up to $934\times$ speedups and is on average $20$ times faster than the state-of-the-art algorithm \algo{SharedMap}~\cite{Schulz25}, with only $12.0\%$ higher communication costs and thereby outperforming \algo{IntMap-S}~\cite{Faraj20}.
    \algo{GPU-IM} delivers extreme speedups (mean $1454\times$, max $12376\times$) at lower solution quality, however it is still comparable the \algo{Fast} configuration of both \algo{SharedMap}~\cite{Schulz25} and \algo{IntMap}~\cite{Faraj20}.
    Future work will focus on improving solution quality without sacrificing performance, e.g., enhancing label propagation filters and refining initial partitioning.

    \bibliography{bib2doi}

\begin{thebibliography}{10}

\bibitem{Acer20}
Seher Acer, Erik~G. Boman, and Sivasankaran Rajamanickam.
\newblock {SPHYNX:} {Spectral Partitioning for HYbrid aNd aXelerator-enabled
  systems}.
\newblock In {\em 2020 {IEEE} International Parallel and Distributed Processing
  Symposium Workshops, {IPDPSW} 2020, New Orleans, LA, USA, May 18-22, 2020}.
  {IEEE}, 2020.
\newblock \href {https://doi.org/10.1109/IPDPSW50202.2020.00082}
  {\path{doi:10.1109/IPDPSW50202.2020.00082}}.

\bibitem{Adams07}
Warren~P. Adams, Monique Guignard, Peter~M. Hahn, and William~L. Hightower.
\newblock {A Level-2 Reformulation-Linearization Technique Bound for the
  Quadratic Assignment Problem}.
\newblock {\em Eur. J. Oper. Res.}, 180(3), 2007.
\newblock \href {https://doi.org/10.1016/j.ejor.2006.03.051}
  {\path{doi:10.1016/j.ejor.2006.03.051}}.

\bibitem{Adams93}
Warren~P. Adams and Terri~A. Johnson.
\newblock {Improved Linear Programming-Based Lower Bounds for the Quadratic
  Assignment Proglem}.
\newblock In Panos~M. Pardalos and Henry Wolkowicz, editors, {\em Quadratic
  Assignment and Related Problems, Proceedings of a {DIMACS} Workshop, New
  Brunswick, New Jersey, USA, May 20-21, 1993}, volume~16 of {\em {DIMACS}
  Series in Discrete Mathematics and Theoretical Computer Science}.
  {DIMACS/AMS}, 1993.
\newblock \href {https://doi.org/10.1090/dimacs/016/02}
  {\path{doi:10.1090/dimacs/016/02}}.

\bibitem{Andreev06}
Konstantin Andreev and Harald R{\"{a}}cke.
\newblock {Balanced Graph Partitioning}.
\newblock {\em Theory Comput. Syst.}, 39(6), 2006.
\newblock \href {https://doi.org/10.1007/S00224-006-1350-7}
  {\path{doi:10.1007/S00224-006-1350-7}}.

\bibitem{Anstreicher03}
Kurt~M. Anstreicher.
\newblock {Recent Advances in the Solution of Quadratic Assignment Problems}.
\newblock {\em Math. Program.}, 97(1-2), 2003.
\newblock \href {https://doi.org/10.1007/s10107-003-0437-z}
  {\path{doi:10.1007/s10107-003-0437-z}}.

\bibitem{Bader14}
David~A. Bader, Henning Meyerhenke, Peter Sanders, Christian Schulz, Andrea
  Kappes, and Dorothea Wagner.
\newblock {Benchmarking for Graph Clustering and Partitioning}.
\newblock In {\em Encyclopedia of Social Network Analysis and Mining}. 2014.
\newblock \href {https://doi.org/10.1007/978-1-4614-6170-8_23}
  {\path{doi:10.1007/978-1-4614-6170-8_23}}.

\bibitem{Bichot13}
C.~E. Bichot and P.~Siarry.
\newblock {\em {Graph Partitioning}}.
\newblock ISTE. Wiley, 2013.
\newblock URL: \url{https://books.google.de/books?id=KUHLscW8D2cC}.

\bibitem{Brandfass12}
B.~Brandfass, T.~Alrutz, and T.~Gerhold.
\newblock {Rank Reordering for Mpi Communication Optimization}.
\newblock {\em Computers \& Fluids}, 80, 2013.
\newblock \href {https://doi.org/10.1016/j.compfluid.2012.01.019}
  {\path{doi:10.1016/j.compfluid.2012.01.019}}.

\bibitem{Buluc16}
Aydin Bulu{\c{c}}, Henning Meyerhenke, Ilya Safro, Peter Sanders, and Christian
  Schulz.
\newblock {R}ecent {A}dvances in {G}raph {P}artitioning.
\newblock In Lasse Kliemann and Peter Sanders, editors, {\em Algorithm
  Engineering - Selected Results and Surveys}, volume 9220 of {\em Lecture
  Notes in Computer Science}. 2016.
\newblock \href {https://doi.org/10.1007/978-3-319-49487-6_4}
  {\path{doi:10.1007/978-3-319-49487-6_4}}.

\bibitem{Burkard98}
Rainer~E. Burkard, Eranda {\c{C}}ela, Panos~M. Pardalos, and Leonidas~S.
  Pitsoulis.
\newblock {\em {The Quadratic Assignment Problem}}.
\newblock Springer US, Boston, MA, 1998.
\newblock \href {https://doi.org/10.1007/978-1-4613-0303-9_27}
  {\path{doi:10.1007/978-1-4613-0303-9_27}}.

\bibitem{Catalyurek23}
{\"{U}}mit~V. {\c{C}}ataly{\"{u}}rek, Karen~D. Devine, Marcelo~Fonseca Faraj,
  Lars Gottesb{\"{u}}ren, Tobias Heuer, Henning Meyerhenke, Peter Sanders,
  Sebastian Schlag, Christian Schulz, Daniel Seemaier, and Dorothea Wagner.
\newblock {M}ore {R}ecent {A}dvances in ({H}yper){G}raph {P}artitioning.
\newblock {\em {ACM} Comput. Surv.}, 55(12), 2023.
\newblock \href {https://doi.org/10.1145/3571808} {\path{doi:10.1145/3571808}}.

\bibitem{Davis11}
Timothy~A. Davis and Yifan Hu.
\newblock {The University of Florida Sparse Matrix Collection}.
\newblock {\em {ACM} Trans. Math. Softw.}, 38(1), 2011.
\newblock \href {https://doi.org/10.1145/2049662.2049663}
  {\path{doi:10.1145/2049662.2049663}}.

\bibitem{Delling09}
Daniel Delling, Peter Sanders, Dominik Schultes, and Dorothea Wagner.
\newblock {Engineering Route Planning Algorithms}.
\newblock In J{\"{u}}rgen Lerner, Dorothea Wagner, and Katharina~Anna Zweig,
  editors, {\em Algorithmics of Large and Complex Networks - Design, Analysis,
  and Simulation {[DFG} priority program 1126]}, volume 5515 of {\em Lecture
  Notes in Computer Science}. Springer, 2009.
\newblock \href {https://doi.org/10.1007/978-3-642-02094-0_7}
  {\path{doi:10.1007/978-3-642-02094-0_7}}.

\bibitem{Dolan02}
Elizabeth~D. Dolan and Jorge~J. Mor{\'{e}}.
\newblock {Benchmarking Optimization Software with Performance Profiles}.
\newblock {\em Math. Program.}, 91(2), 2002.
\newblock \href {https://doi.org/10.1007/s101070100263}
  {\path{doi:10.1007/s101070100263}}.

\bibitem{Edwards14}
H.~Carter Edwards, Christian~R. Trott, and Daniel Sunderland.
\newblock {Kokkos: Enabling Manycore Performance Portability through
  Polymorphic Memory Access Patterns}.
\newblock {\em J. Parallel Distributed Comput.}, 74(12), 2014.
\newblock \href {https://doi.org/10.1016/j.jpdc.2014.07.003}
  {\path{doi:10.1016/j.jpdc.2014.07.003}}.

\bibitem{Faraj20}
Marcelo~Fonseca Faraj, Alexander van~der Grinten, Henning Meyerhenke,
  Jesper~Larsson Tr\"{a}ff, and Christian Schulz.
\newblock {High-Quality Hierarchical Process Mapping}.
\newblock In Simone Faro and Domenico Cantone, editors, {\em 18th International
  Symposium on Experimental Algorithms (SEA 2020)}, volume 160 of {\em Leibniz
  International Proceedings in Informatics (LIPIcs)}, Dagstuhl, Germany, 2020.
  Schloss Dagstuhl -- Leibniz-Zentrum f{\"u}r Informatik.
\newblock \href {https://doi.org/10.4230/LIPIcs.SEA.2020.4}
  {\path{doi:10.4230/LIPIcs.SEA.2020.4}}.

\bibitem{Fiduccia82}
Charles~M. Fiduccia and Robert~M. Mattheyses.
\newblock {A Linear-Time Heuristic for Improving Network Partitions}.
\newblock In James~S. Crabbe, Charles~E. Radke, and Hillel Ofek, editors, {\em
  Proceedings of the 19th Design Automation Conference, {DAC} '82, Las Vegas,
  Nevada, USA, June 14-16, 1982}. {ACM/IEEE}, 1982.
\newblock \href {https://doi.org/10.1145/800263.809204}
  {\path{doi:10.1145/800263.809204}}.

\bibitem{Garey76}
M.~R. Garey, D.~S. Johnson, and L.~Stockmeyer.
\newblock {Some Simplified Np-Complete Graph Problems}.
\newblock {\em Theoretical Computer Science}, 1(3), 1976.
\newblock \href {https://doi.org/10.1016/0304-3975(76)90059-1}
  {\path{doi:10.1016/0304-3975(76)90059-1}}.

\bibitem{Gilbert24}
Michael~S. Gilbert, Kamesh Madduri, Erik~G. Boman, and Siva Rajamanickam.
\newblock {Jet: Multilevel Graph Partitioning on Graphics Processing Units}.
\newblock {\em {SIAM} J. Sci. Comput.}, 46(5), 2024.
\newblock \href {https://doi.org/10.1137/23m1559129}
  {\path{doi:10.1137/23m1559129}}.

\bibitem{Goodarzi19}
Bahareh Goodarzi, Farzad Khorasani, Vivek Sarkar, and Dhrubajyoti Goswami.
\newblock {High Performance Multilevel Graph Partitioning on GPU}.
\newblock In {\em 17th International Conference on High Performance Computing
  {\&} Simulation, {HPCS} 2019, Dublin, Ireland, July 15-19, 2019}. {IEEE},
  2019.
\newblock \href {https://doi.org/10.1109/HPCS48598.2019.9188120}
  {\path{doi:10.1109/HPCS48598.2019.9188120}}.

\bibitem{Gottesbueren20}
Lars Gottesb{\"{u}}ren, Tobias Heuer, Peter Sanders, and Sebastian Schlag.
\newblock {Scalable Shared-Memory Hypergraph Partitioning}.
\newblock In Martin Farach{-}Colton and Sabine Storandt, editors, {\em
  Proceedings of the Symposium on Algorithm Engineering and Experiments,
  {ALENEX} 2021, Virtual Conference, January 10-11, 2021}. {SIAM}, 2021.
\newblock \href {https://doi.org/10.1137/1.9781611976472.2}
  {\path{doi:10.1137/1.9781611976472.2}}.

\bibitem{Hahn12}
Peter~M. Hahn, Yi{-}Rong Zhu, Monique Guignard, William~L. Hightower, and
  Matthew~J. Saltzman.
\newblock {A Level-3 Reformulation-Linearization Technique-Based Bound for the
  Quadratic Assignment Problem}.
\newblock {\em {INFORMS} J. Comput.}, 24(2), 2012.
\newblock \href {https://doi.org/10.1287/ijoc.1110.0450}
  {\path{doi:10.1287/ijoc.1110.0450}}.

\bibitem{Hatazaki98}
Takao Hatazaki.
\newblock {Rank Reordering Strategy for MPI Topology Creation Functions}.
\newblock In Vassil Alexandrov and Jack~J. Dongarra, editors, {\em Recent
  Advances in Parallel Virtual Machine and Message Passing Interface, 5th
  European {PVM/MPI} Users' Group Meeting, Liverpool, UK, September 7-9, 1998,
  Proceedings}, volume 1497 of {\em Lecture Notes in Computer Science}.
  Springer, 1998.
\newblock \href {https://doi.org/10.1007/BFb0056575}
  {\path{doi:10.1007/BFb0056575}}.

\bibitem{Heider72}
Charles~H. Heider.
\newblock {A Computationally Simplified Pair-Exchange Algorithm for the
  Quadratic Assignment Problem}.
\newblock 1972.

\bibitem{Heuer23}
Tobias Heuer.
\newblock {A Direct k-way Hypergraph Partitioning Algorithm for Optimizing the
  Steiner Tree Metric}.
\newblock In Rezaul Chowdhury and Solon~P. Pissis, editors, {\em Proceedings of
  the Symposium on Algorithm Engineering and Experiments, {ALENEX} 2024,
  Alexandria, VA, USA, January 7-8, 2024}. {SIAM}, 2024.
\newblock \href {https://doi.org/10.1137/1.9781611977929.2}
  {\path{doi:10.1137/1.9781611977929.2}}.

\bibitem{Hoefler13}
Torsten Hoefler, Emmanuel Jeannot, and Guillaume Mercier.
\newblock {\em {An Overview of Process Mapping Techniques and Algorithms in
  High-Performance Computing}}.
\newblock 2014.

\bibitem{Holtgrewe10}
Manuel Holtgrewe, Peter Sanders, and Christian Schulz.
\newblock {Engineering a Scalable High Quality Graph Partitioner}.
\newblock In {\em 24th {IEEE} International Symposium on Parallel and
  Distributed Processing, {IPDPS} 2010, Atlanta, Georgia, USA, 19-23 April 2010
  - Conference Proceedings}. {IEEE}, 2010.
\newblock \href {https://doi.org/10.1109/IPDPS.2010.5470485}
  {\path{doi:10.1109/IPDPS.2010.5470485}}.

\bibitem{Karypis98}
George Karypis and Vipin Kumar.
\newblock {A Fast and High Quality Multilevel Scheme for Partitioning Irregular
  Graphs}.
\newblock {\em SIAM Journal on Scientific Computing}, 20(1), 1998.
\newblock \href {https://doi.org/10.1137/S1064827595287997}
  {\path{doi:10.1137/S1064827595287997}}.

\bibitem{Lasalle13}
Dominique Lasalle and George Karypis.
\newblock {Multi-Threaded Graph Partitioning}.
\newblock In {\em 27th {IEEE} International Symposium on Parallel and
  Distributed Processing, {IPDPS} 2013}. {IEEE} Computer Society, 2013.
\newblock \href {https://doi.org/10.1109/IPDPS.2013.50}
  {\path{doi:10.1109/IPDPS.2013.50}}.

\bibitem{LaSalle15}
Dominique LaSalle, Md. Mostofa~Ali Patwary, Nadathur Satish, Narayanan
  Sundaram, Pradeep Dubey, and George Karypis.
\newblock {Improving Graph Partitioning for Modern Graphs and Architectures}.
\newblock In Antonino Tumeo, John Feo, and Oreste Villa, editors, {\em
  Proceedings of the 5th Workshop on Irregular Applications - Architectures and
  Algorithms, {IA3} 2015, Austin, Texas, USA, November 15, 2015}. {ACM}, 2015.
\newblock \href {https://doi.org/10.1145/2833179.2833188}
  {\path{doi:10.1145/2833179.2833188}}.

\bibitem{Lawler63}
Eugene~L. Lawler.
\newblock {The Quadratic Assignment Problem}.
\newblock {\em Management Science}, 9(4), 1963.

\bibitem{Lee24}
Wan{-}Luan Lee, Dian{-}Lun Lin, Tsung{-}Wei Huang, Shui Jiang, Tsung{-}Yi Ho,
  Yibo Lin, and Bei Yu.
\newblock {G-Kway: Multilevel Gpu-Accelerated K-Way Graph Partitioner}.
\newblock In Vivek De, editor, {\em Proceedings of the 61st {ACM/IEEE} Design
  Automation Conference, {DAC} 2024, San Francisco, CA, USA, June 23-27, 2024}.
  {ACM}, 2024.
\newblock \href {https://doi.org/10.1145/3649329.3656238}
  {\path{doi:10.1145/3649329.3656238}}.

\bibitem{Loiola07}
Eliane~Maria Loiola, Nair Maria~Maia de~Abreu, Paulo Oswaldo~Boaventura Netto,
  Peter Hahn, and Tania~Maia Querido.
\newblock {A Survey for the Quadratic Assignment Problem}.
\newblock {\em Eur. J. Oper. Res.}, 176(2), 2007.
\newblock \href {https://doi.org/10.1016/j.ejor.2005.09.032}
  {\path{doi:10.1016/j.ejor.2005.09.032}}.

\bibitem{Manwade18}
K.~B. Manwade and D.~B. Kulkarni.
\newblock {Clustmap: A Topology-Aware Mpi Process Placement Algorithm for
  Multi-Core Clusters}.
\newblock In Subhash Bhalla, Vikrant Bhateja, Anjali~A. Chandavale, Anil~S.
  Hiwale, and Suresh~Chandra Satapathy, editors, {\em Intelligent Computing and
  Information and Communication}, Singapore, 2018. Springer Singapore.

\bibitem{Maue07}
Jens Maue and Peter Sanders.
\newblock {E}ngineering {A}lgorithms for {A}pproximate {W}eighted {M}atching.
\newblock In Camil Demetrescu, editor, {\em Experimental Algorithms, 6th
  International Workshop, {WEA} 2007, Rome, Italy, June 6-8, 2007,
  Proceedings}, volume 4525 of {\em Lecture Notes in Computer Science}.
  Springer, 2007.
\newblock \href {https://doi.org/10.1007/978-3-540-72845-0_19}
  {\path{doi:10.1007/978-3-540-72845-0_19}}.

\bibitem{MullerMerbach70}
Heier M{\"{u}}ller-Merbach.
\newblock {\em {Optimale Reihenfolgen}}.
\newblock Springer-Verlag, 1970.

\bibitem{Pellegrini96}
Fran{\c{c}}ois Pellegrini and Jean Roman.
\newblock {Scotch: A Software Package for Static Mapping by Dual Recursive
  Bipartitioning of Process and Architecture Graphs}.
\newblock In Heather Liddell, Adrian Colbrook, Bob Hertzberger, and Peter
  Sloot, editors, {\em High-Performance Computing and Networking}, Berlin,
  Heidelberg, 1996. Springer Berlin Heidelberg.
\newblock \href {https://doi.org/10.1007/3-540-61142-8_588}
  {\path{doi:10.1007/3-540-61142-8_588}}.

\bibitem{Pellegrini13}
François Pellegrini.
\newblock {\em {Static Mapping of Process Graphs}}, chapter~5.
\newblock John Wiley and Sons, Ltd, 2013.
\newblock \href {https://doi.org/10.1002/9781118601181.ch5}
  {\path{doi:10.1002/9781118601181.ch5}}.

\bibitem{Raghavan07}
Usha~Nandini Raghavan, R\'eka Albert, and Soundar Kumara.
\newblock {Near Linear Time Algorithm to Detect Community Structures in
  Large-Scale Networks}.
\newblock {\em Phys. Rev. E}, 76, 2007.
\newblock \href {https://doi.org/10.1103/PhysRevE.76.036106}
  {\path{doi:10.1103/PhysRevE.76.036106}}.

\bibitem{Sahni76}
Sartaj Sahni and Teofilo Gonzalez.
\newblock {P-Complete Approximation Problems}.
\newblock {\em J. ACM}, 23(3), 1976.
\newblock \href {https://doi.org/10.1145/321958.321975}
  {\path{doi:10.1145/321958.321975}}.

\bibitem{Sanders11}
Peter Sanders and Christian Schulz.
\newblock {E}ngineering {M}ultilevel {G}raph {P}artitioning {A}lgorithms.
\newblock In Camil Demetrescu and Magn{\'{u}}s~M. Halld{\'{o}}rsson, editors,
  {\em Algorithms - {ESA} 2011 - 19th Annual European Symposium,
  Saarbr{\"{u}}cken, Germany, September 5-9, 2011. Proceedings}, volume 6942 of
  {\em Lecture Notes in Computer Science}. Springer, 2011.
\newblock \href {https://doi.org/10.1007/978-3-642-23719-5_40}
  {\path{doi:10.1007/978-3-642-23719-5_40}}.

\bibitem{Schlag20}
Sebastian Schlag, Tobias Heuer, Lars Gottesb{\"{u}}ren, Yaroslav Akhremtsev,
  Christian Schulz, and Peter Sanders.
\newblock {H}igh-{Q}uality {H}ypergraph {P}artitioning.
\newblock {\em CoRR}, abs/2106.08696, 2021.
\newblock \href {https://arxiv.org/abs/2106.08696} {\path{arXiv:2106.08696}},
  \href {https://doi.org/10.48550/arXiv.2106.08696}
  {\path{doi:10.48550/arXiv.2106.08696}}.

\bibitem{Schloegel03}
Kirk Schloegel, George Karypis, and Vipin Kumar.
\newblock {\em {Graph Partitioning for High-Performance Scientific
  Simulations}}.
\newblock San Francisco, CA, USA, 2003.

\bibitem{Schulz17}
Christian Schulz and Jesper~Larsson Tr{\"{a}}ff.
\newblock {Better Process Mapping and Sparse Quadratic Assignment}.
\newblock In Costas~S. Iliopoulos, Solon~P. Pissis, Simon~J. Puglisi, and
  Rajeev Raman, editors, {\em 16th International Symposium on Experimental
  Algorithms, {SEA} 2017, June 21-23, 2017, London, {UK}}, volume~75 of {\em
  LIPIcs}. Schloss Dagstuhl - Leibniz-Zentrum f{\"{u}}r Informatik, 2017.
\newblock \href {https://doi.org/10.4230/LIPIcs.SEA.2017.4}
  {\path{doi:10.4230/LIPIcs.SEA.2017.4}}.

\bibitem{Schulz25}
Christian Schulz and Henning Woydt.
\newblock {S}hared-{M}emory {H}ierarchical {P}rocess {M}apping.
\newblock In {\em Proceedings of the Conference on Applied and Computational
  Discrete Algorithms}. SIAM, 2025.
\newblock \href {https://doi.org/10.1137/1.9781611978759.2}
  {\path{doi:10.1137/1.9781611978759.2}}.

\bibitem{Soper04}
Alan~J. Soper, Chris Walshaw, and Mark Cross.
\newblock {A Combined Evolutionary Search and Multilevel Optimisation Approach
  to Graph-Partitioning}.
\newblock {\em J. Glob. Optim.}, 29(2), 2004.
\newblock \href {https://doi.org/10.1023/B:JOGO.0000042115.44455.f3}
  {\path{doi:10.1023/B:JOGO.0000042115.44455.f3}}.

\bibitem{Traff02}
Jesper~Larsson Tr{\"{a}}ff.
\newblock {Implementing the MPI Process Topology Mechanism}.
\newblock In Roscoe~C. Giles, Daniel~A. Reed, and Kathryn Kelley, editors, {\em
  Proceedings of the 2002 {ACM/IEEE} conference on Supercomputing, Baltimore,
  Maryland, USA, November 16-22, 2002, {CD-ROM}}. {IEEE} Computer Society,
  2002.
\newblock \href {https://doi.org/10.1109/SC.2002.10045}
  {\path{doi:10.1109/SC.2002.10045}}.

\bibitem{Kirchbach20}
Konrad von Kirchbach, Christian Schulz, and Jesper~Larsson Tr{\"{a}}ff.
\newblock {Better Process Mapping and Sparse Quadratic Assignment}.
\newblock {\em {ACM} J. Exp. Algorithmics}, 25, 2020.
\newblock \href {https://doi.org/10.1145/3409667} {\path{doi:10.1145/3409667}}.

\bibitem{Walshaw00}
C.~Walshaw and M.~Cross.
\newblock {Mesh Partitioning: A Multilevel Balancing and Refinement Algorithm}.
\newblock {\em SIAM Journal on Scientific Computing}, 22(1), 2000.
\newblock \href {https://doi.org/10.1137/S1064827598337373}
  {\path{doi:10.1137/S1064827598337373}}.

\bibitem{Yu06}
Hao Yu, I{-}Hsin Chung, and Jos{\'{e}}~E. Moreira.
\newblock {Blue Gene System Software - Topology Mapping for Blue Gene/l
  Supercomputer}.
\newblock In {\em Proceedings of the {ACM/IEEE} {SC2006} Conference on High
  Performance Networking and Computing, November 11-17, 2006, Tampa, FL,
  {USA}}. {ACM} Press, 2006.
\newblock \href {https://doi.org/10.1145/1188455.1188576}
  {\path{doi:10.1145/1188455.1188576}}.

\end{thebibliography}

\end{document}